
\nopagenumbers
\magnification=\magstep1
\parindent=0pt
\font\bigbf=cmbx10
\null
\vskip1truecm
\centerline{\bigbf BLOCK DIAGONALIZING ULTRAMETRIC MATRICES}
\vskip3truecm
\centerline{{\bf T. Temesv\'ari}\footnote*{E-mail:temtam@hal9000.elte.hu}}
\medskip
\centerline{Institute for Theoretical Physics, E\"otv\"os
University,}
\centerline{H-1088 Puskin u 5-7, Budapest, Hungary}
\bigskip
\centerline{\bf C. De Dominicis}
\medskip
\centerline{Service de Physique Th\'eorique, CEA, Saclay}
\centerline{F-91191 Gif-sur-Yvette Cedex, France}
\bigskip
\centerline{\bf I. Kondor}
\medskip
\centerline{Bolyai College, E\"otv\"os University,}
\centerline{H-1145 Amerikai \'ut 96, Budapest, Hungary}
\vskip 3.5truecm
{\bf Abstract}
\medskip The problem of diagonalizing a class of complicated
matrices, to be called ultrametric matrices, is investigated.
These matrices appear at various stages in the description of
disordered systems with many equilibrium phases by the technique
of replica symmetry breaking. The residual symmetry, remaining
after the breaking of permutation symmetry between replicas,
allows us to bring all ultrametric matrices to a block diagonal
form by a common similarity transformation. A large number of
these blocks are, in fact, of size $1\times 1$, i.e.~in a
vast sector the
transformation actually diagonalizes  the  matrix. In  the
other  sectors  we  end  up  with  blocks  of  size $(R+1)\times
(R+1)$ where $R$ is the number of replica symmetry breaking steps.
These blocks cannot be further reduced without giving more
information, in addition to ultrametric symmetry, about the
matrix. Similar results for the inverse of a generic ultrametric
matrix are also derived.

PACS classification numbers: 75.10.Nr, 05.50.+q.

\input epsf
\magnification=\magstep1
\parindent=0.8truecm

\def\a{\alpha}\def\b{\beta}\def\c{\gamma}\def\d{\delta}
\def\q#1#2{q_{#1#2}}
\def\Sum#1{\sum\limits_{#1}}
\def\bra#1{\langle #1\vert}
\def\ket#1{\vert #1\rangle}
\def\Kr#1#2{\d^{\rm Kr}_{#1,#2}}
\def\M#1#2#3#4{M^{#1,#2}_{#3,#4}}
\def\F#1#2{K_k(#1,#2)}
\def\Dk#1{\Delta^{(k)}_{#1}} 
\def\mureg#1#2#3{\mu_{\rm reg}(#1;#2,#3)}
\def\Dd#1#2{{\Delta^{(#1)}_{#2}\over \d_{#2}}}
\null\bigskip\medskip
\noindent{\bf 1. Introduction}
\bigskip

Low temperature disordered systems often possess many
equilibrium phases. The technique of replica symmetry breaking
(RSB) provides a theoretical framework in which these systems can be
described analytically , starting from a microscopic
basis. Discovered and developed in the theory of spin glasses
[1], RSB has
recently penetrated into a number of other problems, including
the theory of random manifolds [2-4],
random field problems [5,6],
protein
folding [7-9],
vortex pinning [10], etc.
In each of these theories randomness is handled via
the replica trick, and the multitude of equilibrium phases is
captured by breaking the permutation symmetry between the
replicas. As always, symmetry breaking means that the low
temperature solutions realize a particular subgroup of the
underlying symmetry group of the theory, here of the permutation
group. The proper choice of the subgroup proved to be a highly
nontrivial task in the case of RSB. The succesful Ansatz for the
symmetry breaking pattern, proposed by Parisi originally in the
context of spin glasses, turned out to embody a particular,
hierarchical organization of the equilibrium phases, usually
referred to as ultrametricity [1].

The corresponding
subgroup determines the structure not only of the order  parameter,
but also of all other quantities in the theory, like
self-energies, propagators, etc. The structure imposed by this
residual symmetry on quantities depending on two replica indices
is by now widely known. The algebra of such quantities has been
worked out by Parisi [11]
with further results, most notably on the inversion
problem, added by M\'ezard and Parisi [2]. At a certain stage of the
development of RSB theories, however, one has to face also more
complicated objects, depending on three or four replica indices.
The structure of these is much harder to grasp and their algebra
is much more involved than that of the two-index quantities. Our
purpose here is to analyse and exploit the structure imposed by
ultametricity on four-index quantities. For reasons to become
clear shortly, we shall call them ultrametric matrices and will
be concerned, in particular, with their diagonalisation and
inversion.

In order to keep the full generality of the results
and thereby guarantee their applicability in any RSB theory, we
shall not assume any properties other than those imposed
by ultrametricity on these matrices. This way we
separate the analysis of  the  purely geometric aspects of
RSB theories, which are common to all of them, from the
treatment of other properties which are determined by more
specific details of the particular systems.

Also, we shall keep
the number $n$ of replicas a positive integer throughout this
paper. The replica limit $n\to 0$ is, of
course, the most essential step of the replica method. It
is also the source of mathematical ambiguities. The analysis of
the consequences of ultrametric symmetry, however, does not depend on
$n$, therefore we found it useful to keep it finite. This way our
analysis belongs to the realm of well-established mathematics
and the analytic continuation in $n$ can be carried out at the latest
stage, on the final results.

The number $R$ of replica symmetry
breaking steps will also be considered a generic integer. Thus
our results will be applicable in situations where only a single
RSB step is needed, as well as in the case of full-fledged RSB
with $R\to \infty$. The results for this "continuous" case ($R\to\infty, n\to
0$), derived by a completely
different method, will be published in [12].

Although almost trivial in principle, our analysis will,
inevitably, be very complicated in actual details. It is clearly
impossible to reproduce the often very lenghty
calculations here, and we shall have to use the phrase "it can be
shown" frequently. What we mean at these points is that one can
reproduce the results easier than to follow the lengthy proofs.
A good strategy is to work out a simple special case (like that
with $R=1$) first; the induction is
easy to spot in most cases.

The plan of the paper is the
following: In Sec.~2 the definition
of ultrametric matrices is presented together with
the classification of their different matrix elements.
Sec.~3 contains a detailed analysis of the non-orthogonal
basis vectors of a similarity transformation that brings
all ultrametric matrices to a block diagonal form.
In Sec.~4 this block diagonal
form is expressed through some
"kernels", which facilitate the eigenvalue and
inversion problem greatly. A complete list of matrix components versus kernels
is also included in this section. Some technicalities are relegated
to the Appendix.
\bigskip\medskip
\noindent{\bf 2. Definition of a generic ultrametric matrix}
\bigskip

For the sake of definiteness we present our analysis in the language
of spin glasses,
the extension to other replica symmetry breaking theories is
merely a  matter of notation. The replica method yields the
free-energy $F$ of a long-range spin glass in the form of a
functional, depending on a set $\q{\a}{\b}$  of order parameters:
$F=F(\q{\a}{\b})$.
The replica indices $\a,\b$ take  {\bf integer}  values:
$\a,\b=1,2,\ldots,n$ (for our present analysis the replica
limit $n\to 0$ need not be considered here). The order
parameters are symmetric: $\q{\a}{\b}=\q{\b}{\a}$, and (for Ising spins) the
diagonal components are zero: $\q{\a}{\a}=0$. The number of independent
order parameter components is thus ${1\over 2} n(n-1)$. The free energy is
independent of the labeling of the replicas, so $F$ must be
constructed from the algebraic invariants of the permutation group
of $n$ objects. Examples of such invariant combinations are:
$$\eqalign{\Sum {\a\b}\q{\a}{\b},\qquad Tr\,q^2=\Sum {\a\b}\q{\a}{\b}^2,
\qquad Tr\,q^3=\Sum {\a\b\c}\q{\a}{\b}\q{\b}{\c}\q{\c}{\a},\cr
\Sum {\a\b}\q{\a}{\b}^4,\qquad Tr\,q^4,\qquad \Sum {\a\b\c}
\q{\a}{\b}^2\q{\b}{\c}^2,\qquad \Sum {\a\b\c\d}
\q{\a}{\b}^2\q{\a}{\c}\q{\c}{\d}\q{\d}{\a},\qquad{\rm etc.}\cr}$$

  The stationary values of the order parameter components are determined
by the equation of state: ${\partial F\over\partial\q{\a}{\b}}=0$.
There are ${1\over 2} n(n-1)$ such equations.
Depending on the parameters in $F$ these equations may have
solutions that are indentically zero, $\q{\a}{\b}=0$, $\forall \a,\b$,
or that have
non-zero but identical off-diagonal components
$\q{\a}{\b}=q(1-\Kr {\a}{\b})$, or non-zero
off-diagonal components that depend on the pair ($\a,\b$) of replica
indices. The solutions of the last kind are said to be replica
symmetry breaking (RSB) solutions and these are the ones that
describe situations where many equilibrium states exist. Led by
a number of formal considerations that later turned out to
embody the ultrametric organization of these states, Parisi
proposed a, by now standard, parametrization for the RSB
solutions which we briefly recapitulate in order to fix
notations.

Firstly assume that $n$ is not only an integer, but a
very large one, with a large number of proper divisors. Let $p_1$ be one
of these, itself a large number with many divisors, one of them
$p_2$, etc.~up to $p_R$. It is useful to rename $n$ as $p_0$, and add
$p_{R+1}\equiv 1$ to the other end of the series.
Now the $n$ replicas are divided into $n/p_1$ boxes each
containing $p_1$ replicas. The contents of each box are
further subdivided into $p_1/p_2$ smaller boxes with $p_2$
replicas in each etc., down to the smallest boxes with $p_R$
replicas. The RSB solutions are supposed to be invariant
w.r.t.~the permutations of replicas inside each of the smallest
boxes of size $p_R$, and also w.r.t.~the permutations of the
size $p_{k+1}$ boxes inside each of the size $p_k$ boxes for any
$k=0,1,\ldots,R-1$. Evidently, these permutations form a
subgroup of the permutation group. This subgroup is the residual
symmetry that remains after the breaking of replica symmetry.
The Ansatz for the order parameter matrix corresponding to this
residual symmetry is constructed as follows: The
$n\times n$ (i.e.~$p_0\times p_0$)
matrix $\q{\a}{\b}$ is divided into blocks of size $p_1\times p_1$,
and a common
value $q_0$ is assigned to all matrix elements outside the
diagonal blocks. Next the diagonal blocks are further divided
into blocks of size $p_2\times p_2$, the value $q_1$ assigned to the matrix
elements inside the diagonal blocks of size $p_1\times p_1$ but outside the
diagonal blocks of size $p_2\times p_2$, etc., down to the innermost blocks
of size $p_R\times p_R$, where the matrix elements are $q_R$ except along the
very diagonal of the whole matrix where they are zero.
Some formulae below (eqs.~32,34) become meaningless whenever the
ratio of two subsequent $p$'s is $2$ or $3$. These cases would
require a separate discussion which we can safely omit here,
since in practical applications these cases will never
appear. For our present purposes we can stipulate
$p_k/p_{k+1} >3,\quad k=0,1,\ldots ,R$.

The solution of
the stationary condition ${\partial F\over\partial\q{\a}{\b}}=0$ is sought
among the matrices which have
the special form just described. This solution is a point in the
${1\over 2}n(n-1)$-dimensional replica space, so it is, in fact,
a {\bf vector}. In
the following, when we deal with genuine matrices
acting on replica space, i.e.~with quantities depending on two
pairs of replica indices, we will actually call $\q{\a}{\b}$ and similar
quantities vectors. The association between the $n\times n$ symmetric
matrix $\q{\a}{\b}$ (with $\q{\a}{\a}=0$) and the column vector
$\ket {\q{\a}{\b}}$ is evident: one
lists the matrix elements above the diagonal of $\q{\a}{\b}$ in any
prescribed order (say, row by row) below each other.

The
representation of $\q{\a}{\b}$ and other vectors of replica space by
symmetric matrices remains, nevertheless, very useful, because
it is much easier to display their special structure in the
matrix form. Therefore, we shall use this matrix representation
for all vectors appearing in this paper. For later reference
we note here that the scalar product of two vectors, $\ket {r_{\a\b}}$ and
$\ket {\q{\a}{\b}}$, is, in matrix language, half the trace of the product
of the corresponding matrices:
$$\bra {r_{\a\b}}\q{\a}{\b}\rangle={1\over 2}\,Tr(rq).\eqno (1)$$

We now introduce the concept of the overlap between replica
indices that will play a central role in the following: the
overlap between $\a$ and $\b$ is $k$ (notation: $\a\cap\b=k$) if in the
Parisi scheme $\q{\a}{\b}=q_k$.  The overlap $\a\cap\b$ defined this way can be
regarded as a kind of hierarchical distance between replicas $\a$
and $\b$, its values ranging from $0$ (corresponding to the
largest off-diagonal blocks of size $p_1\times p_1$) to $R+1$ (corresponding
to the diagonal, $\a=\b$).

It is evident that any quantity $f$
constructed of the $q$'s and depending on only two replica indices
(such as $f_{\a\b}=\Sum {\c}\q{\a}{\c}\q{\c}{\b}$, for example)
depends only on their overlap: $f_{\a\b}=f(\a\cap\b)$.

The metric generated by the overlaps is, by construction,
ultrametric: whichever way we choose three replicas $\a,\b,\c$, either
all three of their overlaps are the same ($\a\cap\b=\a\cap\c=\b\cap\c$),
or one (say $\a\cap\b$)
is larger than the other two, but then these are equal ($\a\cap\b>
\a\cap\c=\b\cap\c$).

Furthermore, it also follows that any quantity $f$ built of the
$q$'s and depending on three replica indices, $f_{\a\b\c}$, depends only on
the overlaps $\a\cap\b$, $\a\cap\c$, $\b\cap\c$, and since of these
at most two can be
different, $f_{\a\b\c}$ is, in fact, a function of only two variables,
e.g.~of $\a\cap\b$ and of the larger of the other two:
$$f_{\a\b\c}=f(\a\cap\b;{\rm max} \{\a\cap\c ,\b\cap\c\}).\eqno(2)$$

In the following we will also have
to consider quantities depending on four replica indices, coming
in two pairs: $f_{\a\b,\c\d}$. A little reflection shows that such a
quantity can always be parametrised as follows:
$$f_{\a\b,\c\d}=f^{\a\cap\b,\c\cap\d}_{{\rm max}\{\a\cap\c,\a\cap\d\},\,
{\rm max}\{\b\cap\c,\b\cap\d\}}.\eqno(3)$$

Admittedly, this parametrisation is less than perfect.
Firstly, ultrametricity implies that of the six possible overlaps
between $\a$, $\b$, $\c$ and $\d$ at most three can be different, which
corresponds to the simple geometric fact that the edges of a
tetrahedron having equilateral or isosceles faces can have at
most three different lengths. Therefore, of the four variables
on the r.h.s.~of (3) at least two are the same. The resulting
redundancy is the price we pay for the symmetry of the notation.
Secondly, in all practical applications $f_{\a\b,\c\d}$ is symmetric
w.r.t.~exchanging $\a$ and $\b$ or $\c$ and $\d$ and also w.r.t.
exchanging the two pairs: $f_{\a\b,\c\d}=f_{\a\b,\d\c}=f_{\b\a,\c\d}=
f_{\c\d,\a\b}$ etc., and these symmetries are not
manifestly reflected by the parametrisation (3). We prefer
keeping the consequences of these symmetries in mind rather than
overcomplicating the notation.

The choice between the various
types of solutions of the equation of state (identically zero,
or constant $\q{\a}{\b}$, or replica symmetry broken $\q{\a}{\b}$)
is based on
stability considerations. In order to decide the stability of a
given solution, one has to diagonalize the Hessian or (bare)
self-energy matrix ${\partial^2F\over\partial \q{\a}{\b}\partial
\q{\c}{\d}}=M_{\a\b,\c\d}$, evaluated at the stationary point. $M$
is the prime example of a quantity depending on two pairs of
replica indices, so it can be parametrised as shown in (3). $M$
is obviously symmetric w.r.t.~the exchange of the two pairs
($\a\b$) and ($\c\d$). Since $\q{\a}{\b}=\q{\b}{\a}$ and $\q{\a}{\a}=0$,
$M$ can be considered to
depend on the ordered pairs $\a <\b$ and $\c <\d$ only, so it is a matrix
of dimension  ${1\over 2}n(n-1)\times {1\over 2}n(n-1)$.
A symmetric matrix of this size has, in
general, ${1\over 4}n(n-1)[{1\over 2}n(n-1)+1]$
independent elements. This number is greatly
reduced by ultrametricity. Below we list all the different
kinds of matrix elements that can appear. When doing so, we will
relax the ordering of the indices of $M_{\a\b,\c\d}$, and extend the
definition to arbitrary combinations of the indices (except $\a=\b$
and $\c=\d$) such as to make $M$ symmetric w.r.t.~exchanging $\a$
and $\b$ and/or $\c$ and $\d$, $M_{\a\b,\c\d}=
M_{\a\b,\d\c}=M_{\b\a,\c\d}=M_{\b\a,\d\c}$, in addition to the symmetry
w.r.t.~exchanging the pairs ($\a\b$), ($\c\d$). This extension is
motivated by convenience: when summations are to be performed on
the indices of $M$ the restrictions due to ordering can
become very cumbersome.

The matrix elements can be classified
naturally in three categories:

(i) \underbar{Matrix elements of the first kind.}
These are the diagonal elements $M_{\a\b,\a\b}$,
together with their variants $M_{\a\b,\b\a}$, $M_{\b\a,\a\b}$,
etc. They depend on the overlap $\a\cap\b=i=0,1,2,\ldots,R$
only. Under the parametrisation (3) they are given by
$$M_{\a\b,\a\b}=M^{i,i}_{R+1,R+1}\quad\quad\quad
i=0,1,\ldots,R.\eqno(4)$$
There are, in general, $R+1$ different matrix elements in this
category (instead of ${1\over 2}n(n-1)$, the dimension of the matrix).

(ii)\underbar{Matrix elements of the second kind.} These are off-diagonal,
with one replica index in common between the two pairs. One
example is $M_{\a\b,\a\c}$ which, together with its exchanged variants
($M_{\a\b,\c\a}$ etc.), exhausts all possibilities.

There are three cases:
\medskip
\item{(a)} $\a\cap\b=\a\cap\c=i\le\b\cap\c=j=0,1,\ldots,R$.
Then
$$M_{\a\b,\a\c}=M^{i,i}_{R+1,j}\quad,\quad\quad j\ge i.\eqno (5)$$
Various exchanges of the replica indices either reproduce
the same, or exchange the lower variables $R+1$ and $j$. (The
parametrisation (3) is such that $M^{i,j}_{k,l}$ is always symmetric in $k$
and $l$.)
\item{(b)} $\a\cap\b=\b\cap\c=i<\a\cap\c=j$.
$$M_{\a\b,\a\c}=M^{i,j}_{R+1,i}\quad,\quad\quad j>i.\eqno (6)$$
Exchanging replica indices in all possible
ways reproduces either the same, or exchanges the lower
variables, or exchanges $i$ and $j$. Thus:
$$M^{i,j}_{R+1,i}=M^{i,j}_{i,R+1}=M^{j,i}_{R+1,j}=M^{j,i}_{j,R+1}\quad,\quad
\quad i<j.\eqno (7)$$
\item{(c)} $\a\cap\b=i>\a\cap\c=\b\cap\c=j$.
$$M_{\a\b,\a\c}=M^{i,j}_{R+1,i}\quad,\quad\quad i>j.$$
According to (7), this is the same as (6) (rename
$i\leftrightarrow j$).

It is
easy to see that the number of different matrix elements
is at most $(R+1)^2$ in this class.

(iii)\underbar{ Matrix elements of the third kind.}
These have
four different replica indices,
\penalty-10000 $M_{\a\b,\c\d}$. Considering all
logically possible situations with $\a<\b$, $\c<\d$,
$\a<\c$, $\b<\d$ (corresponding to the
matrix elements above the diagonal of $M$), we find
six possible cases altogether.
\medskip
\item{(a)} $\a\cap\b=i$, $\c\cap\d=j$, ${\rm max}\{\a\cap\c,\a\cap\d\}=
{\rm max}\{\b\cap\c,\b\cap\d\}=k$ with $k\le {\rm min}\{i,j\}$.
Then
$$M_{\a\b,\c\d}=M^{i,j}_{k,k}=M^{j,i}_{k,k},\eqno (8)$$
where the second equality follows from exchanging
the two pairs $(\a\b)\leftrightarrow (\c\d)$.
\item{(b)} $\a\cap\b=i$, $\c\cap\d=j$, ${\rm
max}\{\a\cap\c,\a\cap\d\} ={\rm max}\{\b\cap\c,\b\cap\d\}=k$,
$j<k\le i$.
Then $$M_{\a\b,\c\d}=M^{i,j}_{k,k}=M^{j,i}_{k,j}.\eqno (9)$$
\item{(c)} $\a\cap\b=i$, $\c\cap\d=j$, ${\rm max}\{\a\cap\c,\a\cap\d\}=i$,
${\rm max}\{\b\cap\c,\b\cap\d\}=k$, $j\le i<k$.
Then
$$M_{\a\b,\c\d}=M^{i,j}_{i,k}=M^{j,i}_{k,j}=M^{j,i}_{j,k}.\eqno (10)$$
\item{(d)} $\a\cap\b=i$, $\c\cap\d=j$, ${\rm max}\{\a\cap\c,\a\cap\d\}=i$,
${\rm max}\{\b\cap\c,\b\cap\d\}=k$, $i<j\le k$.
Then $$M_{\a\b,\c\d}=M^{i,j}_{i,k}=M^{j,i}_{k,j}.\eqno (11)$$
\item{(e)} $\a\cap\b=i$, $\c\cap\d=j$, ${\rm max}\{\a\cap\c,\a\cap\d\}=i$,
${\rm max}\{\b\cap\c,\b\cap\d\}=k$, $i<k<j$.Then
$$M_{\a\b,\c\d}=M^{i,j}_{i,k}=M^{j,i}_{k,k}.\eqno (12)$$
\item{(f)} $\a\cap\b=\c\cap\d=i$, ${\rm max}\{\a\cap\c,\a\cap\d\}=k$,
${\rm max}\{\b\cap\c,\b\cap\d\}=l$, $i<{\rm min}\{k,l\}$.
Then
$$M_{\a\b,\c\d}=M^{i,i}_{k,l}.\eqno (13)$$

\noindent In eqs.~(8)-(13) the overlaps $i$,$j$,$k$,$l$ can run through
$0,1,2,\ldots ,R$. Considering the cases (a)-(f) one can show
that the number of different matrix elements of the third kind is $(R+1)^3$.

If all the ${1\over 2}n(n-1)$ independent order parameter components were
different, the matrix $M_{\a\b,\c\d}$ would have of order $n^4$ independent
matrix elements. Parisi's RSB scheme does not completely destroy
the permutation symmetry of the replicas, however, it only
reduces this symmetry to a particular subgroup of the group of
permutations of $n$ elements. It is this residual symmetry that
is responsible for the tremendous reduction in the number of
independent elements of the Hessian: instead of $O(n^4)$ we have,
according to eqs.~(4)-(13), only $O(R^3)$ different matrix elements,
which, for large $n$, is exponentially small compared to $n^4$.

The particular structure described above has been displayed in
the example of the Hessian of the long-range spin glass.
Matrices with an identical structure appear in many RSB
theories. We shall call these matrices ultrametric matrices.
{}From this point on we shall disregard the derivation and meaning
of $M$, and will focus solely on its symmetries. It will be seen
that these symmetries allow one to construct an irreducible
representation for ultrametric matrices in that all those
that have the same block sizes $p_0,p_1,\ldots ,p_R$ can
be brought to a block diagonal form by the same similarity
transformation and that no further reduction is possible without
providing further information on the matrix elements. It will
also be seen that the conditions ultrametricity imposes upon $M$
are stringent enough to actually yield a large number of the
eigenvalues in closed form. We shall also look into the problem
of inversion of ultrametric matrices and shall find again that a
large number of the components of the inverse can be obtained in
closed form.  Some of the results we compile here are not new,
they were published by two of us some years ago in a very
compact form [13]. In addition to rephrasing them and providing some
background material we also present a number of new results,
especially with regards to the inversion problem.  As long as $n$ is an
integer with the sequence of divisors $p_1,p_2,\ldots,p_R$ as described,
the matrix $M$ is a
well-defined mathematical object, and the problem of its
diagonalisation belongs to the realm of standard mathematics.
The present paper will be concerned with this well-posed
problem. At a certain point one will, however, have to consider
the replica limit $n\to 0$, together with the analytic continuation
in all the $p_i$'s and with the limit $R\to \infty$, as proposed
by Parisi [1].
These manipulations are at the present time of a purely formal
character, certainly beyond the limits of well-established
mathematics. After all these dubious steps one arrives at the problem
of the diagonalisation of an integral operator with a set of
particular symmetries. The results we get in the discrete case
can all be easily transcribed on to this new, continuous
problem.
\goodbreak
\bigskip\medskip
\noindent{\bf 3. The new basis}
\bigskip
In the previous section the components of an ultrametric matrix
$M_{\a\b,\c\d}$ have been given in the Cartesian coordinate system spanned
by the basis vectors $\ket {\mu,\nu}$, $(\mu,\nu)=(1,2),(1,3),\ldots,(n-1,n)$,
which, similarly to the order
parameter $\q{\a}{\b}$, can be represented by symmetric $n\times
n$ matrices. Their
matrix
elements are
$$\ket{\mu,\nu}_{\a\b}=\d^{{\rm Kr}}_{(\mu,\nu),(\a,\b)}=
\cases{1,&if $\mu=\a,\nu=\b$ or $\mu=\b,\nu=\a$;\cr
0,&otherwise.\cr}\eqno (14)$$

In order to bring $M$ to a block diagonal form, we have to
go over into a new basis. The new set of basis vectors can be
inferred from the general structure of
the eigenvectors described in [13] and, like those,
can be naturally classified in three families.
\medskip
\underbar{The first family}

The first family of basis vectors consists of $R+1$ vectors
labelled by $i=0,1,\ldots,R$ which, when represented by quadratic
matrices, have identical nonzero elements on the $i^{\rm th}$ level of
the Parisi hierarchy and zeros everywhere else:
$$\ket{0;i}=\left({1\over 2}n(p_i-p_{i+1})\right)^{-{1\over2}}
\sum_{\a\cap\b=i}\ket{\a,\b}\eqno (15)$$
where the $\ket{\a,\b}$'s are the Cartesian unit vectors defined in
(14). The meaning of the first label ($0$) will become clear shortly.

The first family basis vectors form an $(R+1)$-dimensional orthonormal set
in replica space. The difference $p_i-p_{i+1}$ in the normalisation factor
will appear so often in the following that it is worth giving it a
name:
$$p_i-p_{i+1}=\d_i\quad,\quad\quad i=0,1,\ldots,R.\eqno (16)$$

A straightforward but tedious
calculation shows that the subspace spanned by the first family
basis vectors is closed under the action of an ultrametric
matrix. Therefore the linear combination
$$\ket{f}=\sum_{i=0}^R\,f_0(i)\,\ket{0;i}\eqno (17)$$
is an
eigenvector of $M$, provided the amplitudes $f_0(i)$ are
appropriately chosen. The conditions for these amplitudes (i.e.~
the eigenvalue equations) will be written up in the next
section. Evidently, there will be $R+1$ possible choices for the
amplitudes, corresponding to $R+1$ eigenvalues $\lambda_m(0),\,m=0,1,\ldots,R$.
In the case of
a generic matrix $M$ with no symmetries other than
those dictated by ultrametricity, all these eigenvalues will be
singlets, their multiplicity $\mu(0)=1$, and the eigenvectors orthogonal.
In the following we shall often refer to the first family as the longitudinal
or L family.
\medskip
\underbar{The second family}

The second family will be broken down into several subfamilies,
to be labelled by an index $k=1,2,\ldots,R+1$. The first family is, in several
respects, nothing but the case corresponding to $k=0$, which is why we
used the label $0$ in addition to $i$. The structure of the
second family basis vectors is easiest to grasp graphically, so
we define them in a series of figures.  The vectors belonging to
the \underbar{$k=1$ subfamily} are shown in Figs.~l,2.
(Although, as we have already mentioned, the ratios of
subsequent $p$'s must never be $2$ or $3$, in order to prevent
the figures from occupying an excessive space, here and in
almost all the figures to follow we have to illustrate the
structure of eigenvectors by figures where some of these ratios are
$3$.) Consider the vectors
shown in Fig.~l. They have nonzero components only on the zeroth
level of the Parisi hierarchy, but now not all these components
are identical: they take two different values, $A$ and $B$,
arranged as shown in the figure. We shall denote these vectors
as $\ket{1;0;b}$, where the first label is the value of $k$, the second is
that level of the Parisi hierarchy where the vector has nonzero
elements, and the third, $b=1,2,\ldots,n/p_1$, shows which column and row of
blocks is distinguished, i.e.~which blocks have matrix elements $B$.
\medskip
\vbox{\centerline{
\epsfxsize=13truecm\epsfbox{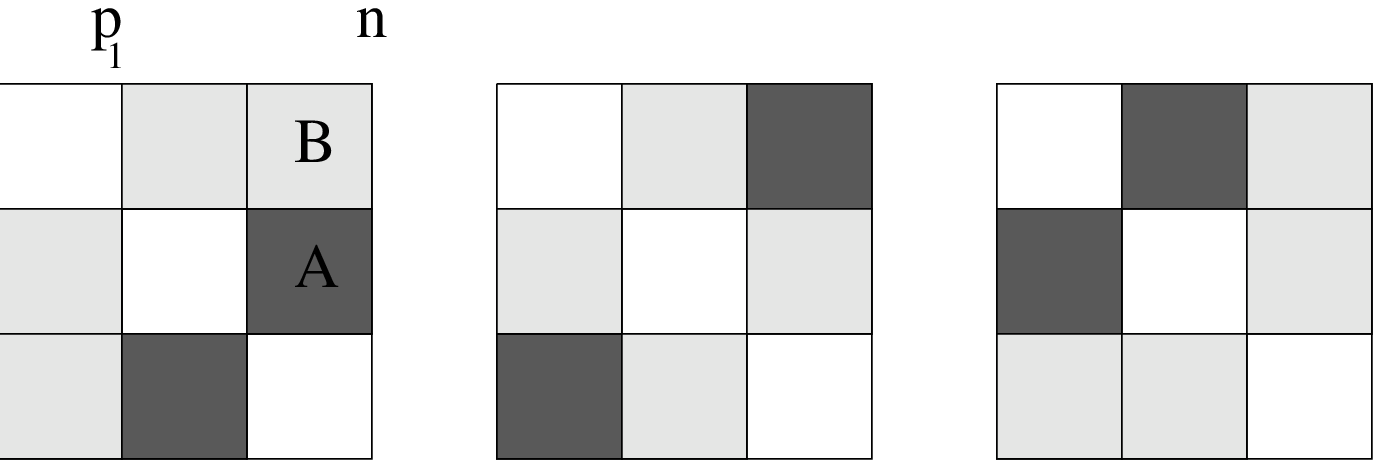}}
\smallskip
\noindent  Fig.~l: The vectors $\ket{k=1,i=0,b}$. The label
$b=1,2,\ldots,n/p_1$
shows which
column and row of $p_1\times p_1$ blocks is distinguished. Identical shading
means identical components. Blank means zero.\par}
\bigskip\medskip
Now consider the sum of these vectors, $\sum^{n/p_1}_{b=1}\ket{1;0;b}$.
The distinction between the different blocks will obviously
disappear in the sum, so it will be proportional to $\ket{0;0}$ of the
first family. However, we want to make each of the second family vectors
orthogonal to the first family, so we have to choose
the vector components, $A$ and $B$ so as to make the above
sum vanish. If we choose, say, $A=1$ then $B$ must be
$$B={1\over2}\left(2-{n\over p_1}\right).\eqno (18)$$
With this choice the $\ket{1;0;b}$ vectors are all orthogonal to the
first family and
$$\sum^{n/p_1}_{b=1}\ket{1;0;b}=0.\eqno (19)$$
The evident symmetry between these vectors makes it obvious
that they pairwise make the same angle which, together with
(19), means that they span an $({n\over p_1}-1)$-dimensional hypertetrahedron.
It also follows that there is no further linear relationship
between them, so if we discard one, say $\ket{1;0;n/p_1}$, we will
be left with
$$\mu(1)=n\left({1\over p_1}-{1\over p_0}\right)\eqno (20)$$
linearly independent basis vectors. These will not be
normalised, nor orthogonal, however. It would be an easy task to
construct an orthonormal set out of them, but it would destroy
their symmetry.  We find it slightly more convenient to work
with a biorthogonal set. For the same reason we need not
worry about normalisation. It is an elementary exercise to show
that the set
$${4p_1\over n^2(n-2p_1)}\left(\ket{1;0;b}-\ket{1;0;{n\over p_1}}\right),
\quad b=1,2,\ldots,{n\over p_1}-1,\eqno (21)$$
is biorthogonal to the set $\ket{1;0;b}$.

We now proceed, still
within the $k=1$ subfamily, to the next level of the Parisi
hierarchy. The vectors $\ket{k=1;i=1;b}$ are shown in Fig.~2.
\medskip
\vbox{\centerline{
\epsfxsize=15truecm\epsfbox{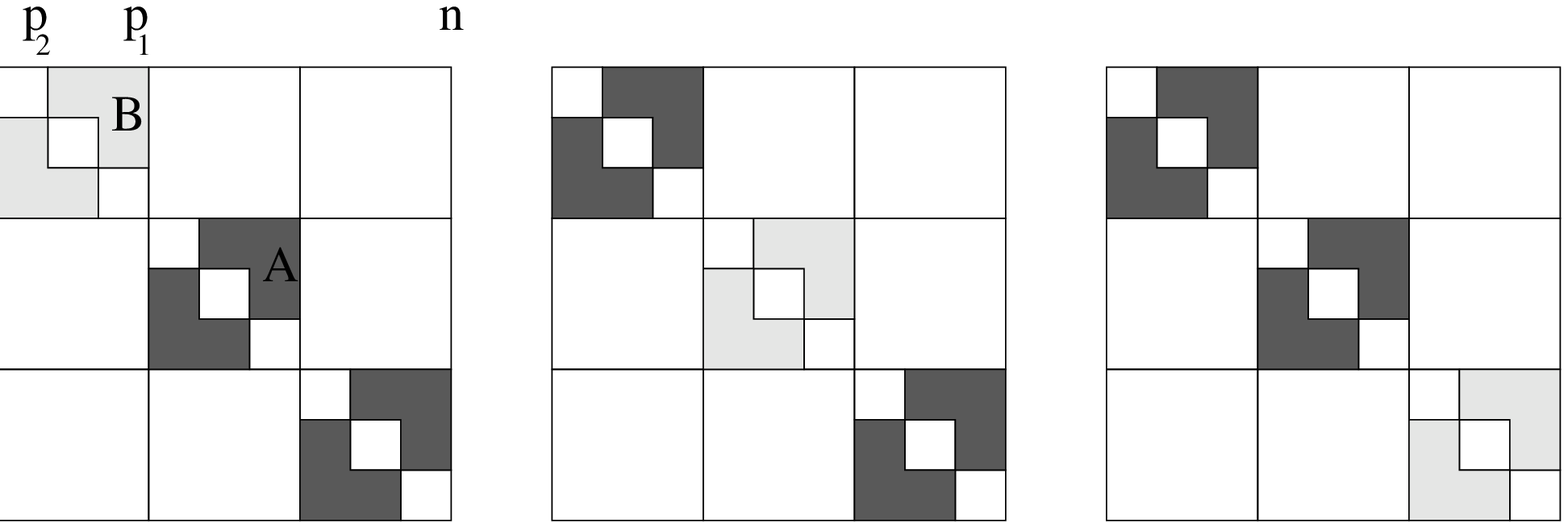}}
\smallskip
\noindent
Fig.~2.: The basis vectors $\ket{1;1;b}$.\par}
\bigskip\medskip
\penalty-10000
Similarly to the previous case, orthogonality to the first
family vector $\ket{0;1}$ demands
$$\sum_{b=1}^{n/p_1}\,\ket{1;1;b}=0$$
which is satisfied if
the vector components are chosen as
$$A=1,\quad\quad B=1-{n\over p_1}.\eqno (22)$$
Then the vectors $\ket{1;1;b}$ span a $\mu(1)$-dimensional
hypertetrahedron again, with the associated biorthogonal set
$${2p_1\over n^2(p_1-p_2)}\left(\ket{1;1;b}-\ket{1;1;{n\over p_1}}\right),
\quad b=1,2,\ldots,{n\over p_1}-1.\eqno (23)$$
The construction proceeds along similar lines: filling in the
$i^{\rm th}$ level of the Parisi hierarchy we find the same
$\mu(1)$-dimensional tetrahedra with the same orthogonality
conditions (22) applying for any $i\ge1$ (only the case $i=0$,
eq.~(18), is
different).  This way we will have, altogether, $\mu(1)\,(R+1)$ independent
basis vectors
$$\ket{1;i;b},\quad\quad i=0,1,\ldots,R\quad {\rm and}\quad b=1,2,\ldots,
{n\over p_1}-1,$$
making up the $k=1$ subfamily. They are each
orthogonal to the first family, two of them belonging to
different values of $i$ are also orthogonal, but two such vectors
with the same $i$ and different $b$'s are not.

It can now be
shown again that the subspace spanned by the $R+1$ vectors $\ket{1;i;b}$,
$i=0,1,\ldots,R$, for $b$ {\bf fixed}
is an invariant subspace of an arbitrary ultrametric
matrix. Therefore the linear combination
$$\ket{f}=\sum_{i=0}^{R}\,f_1(i)\,\ket{1;i;b}$$\goodbreak
\noindent with
appropriately chosen amplitudes $f_1(i)$, independent of $b$, will be
an eigenvector. There will be $R+1$ choices for these amplitudes,
yielding $R+1$ eigenvalues $\lambda_1(m)$, $m=0,1,\ldots,R$,
in the $k=1$ subfamily. Each of
these will be $\mu(1)$-fold degenerate, according to the free choice
of $b$.

\penalty-10000
We now turn to the \underbar{$k=2$ subfamily}. Some of the $k=2$ type
vectors with the $i=0$ Parisi level filled are shown in Fig.~3.
\medskip
\vbox{\centerline{
\epsfxsize=14truecm\epsfbox{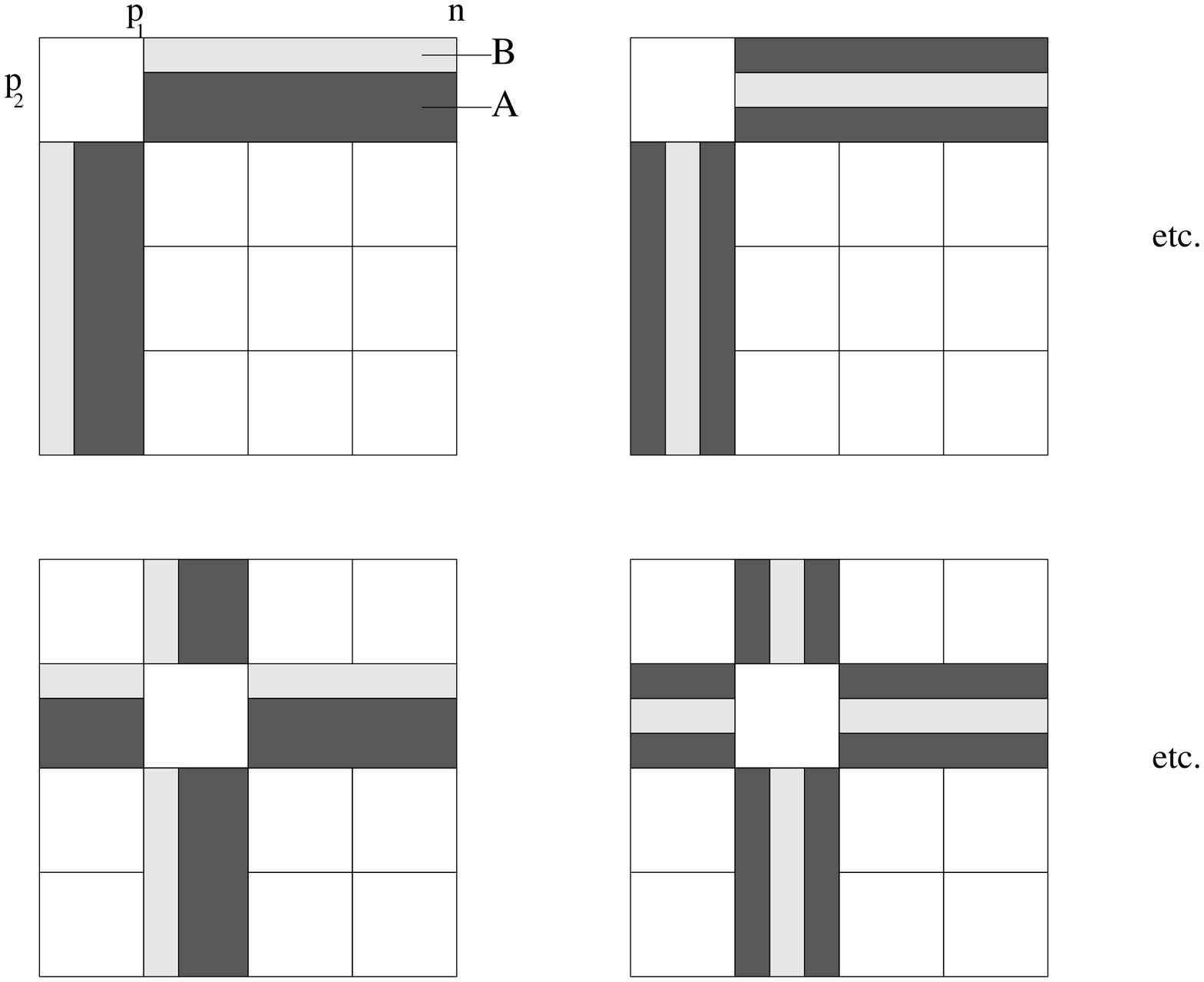}}
\smallskip
\noindent
Fig.~3: Some vectors of the $k=2$ subfamily $(i=0)$.\par}
\bigskip
These vectors will be labelled as $\ket{2;0;a,b}$ where $a=1,2,\ldots,n/p_1$
shows which column of $p_1\times p_1$ sized blocks has nonzero elements and
$b=1,2,\ldots,p_1/p_2$ shows which column of $p_2\times p_2$
sized blocks is distinguished
inside column $a$. The sum $\sum^{p_1/p_2}_{b=1}\ket{2;0;a,b}$
must vanish again for any fixed
$a$, otherwise it would be a linear combination of $k=0$ and $k=1$
subfamily type vectors.
This
orthogonality condition demands that we choose
$$A=1,\quad\quad B=1-{p_1\over p_2}\eqno (24)$$
leaving $p_1/p_2-1$ independent vectors (spanning a tetrahedron
again) for any fixed $a$. It is easy to see that with this
choice the vectors $\ket{2;0;a,b}$ will not only be orthogonal to each of the
previous families (with $k=0,1$) but they will also be orthogonal to
the vectors $\ket{2;0;a',b'}$ with $a\ne a'$ and any $b'$.

Some $k=2$, $i=1$ vectors are
shown in Fig.~4.
\medskip
\vbox{\centerline{
\epsfxsize=10truecm\epsfbox{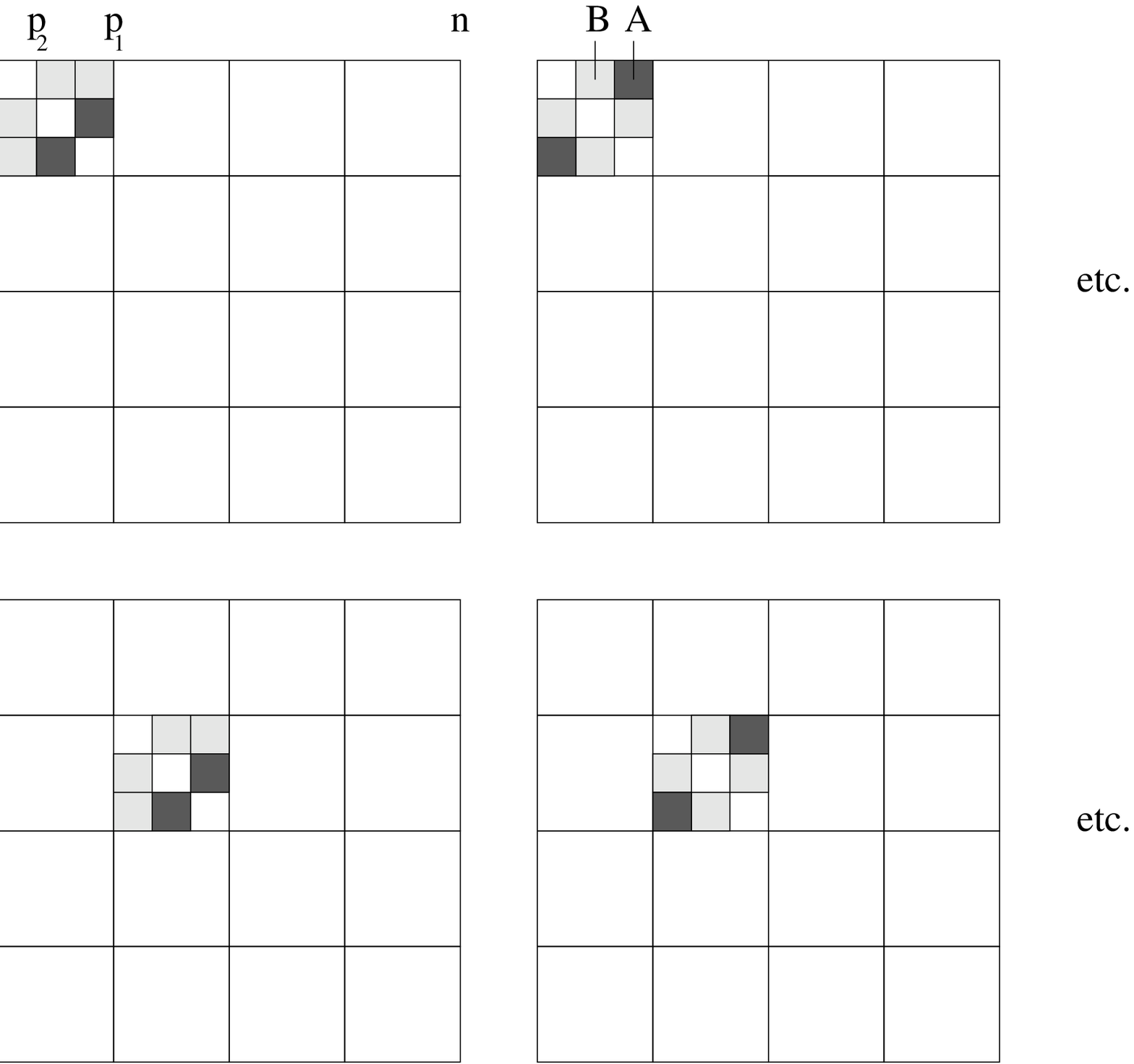}}
\smallskip
\noindent
Fig.~4.: Some vectors of the $k=2$ subfamily $(i=1)$.\par}
\bigskip\medskip
The orthogonality conditions now read
$$A=1,\quad\quad B={1\over 2}\left(2-{p_1\over p_2}\right).$$
We can go on to build $\ket{2;i;a,b}$, $i=0,1,\ldots,R$
in a similar manner. The subspace of
these vectors for fixed $a$ and $b$ will be closed under the
action of an ultrametric matrix $M$. Thus
$$\sum^{R}_{i=0}\,f_2(i)\,\ket{2;i;a,b}$$
will be an
eigenvector with $R+1$ choices for the amplitudes and with
eigenvalues $\lambda _2(m)$, $m=0,1,\ldots,R$.
We have seen that for a given $a$ we have ${p_1\over p_2}-1$
linearly independent choices for the basis vectors, while for
different $a$'s they are already orthogonal. That means we have
${n\over p_1}\left({p_1\over p_2}-1\right)$, i.e.
$$\mu(2)=n\left({1\over p_2}-{1\over p_1}\right)$$
independent basis vectors for any $i$. The
total dimension of the $k=2$ subfamily is thus $\mu(2)\,(R+1)$, and the
multiplicity of the $k=2$ eigenvalues is $\mu(2)$.

The generalisation
is now obvious. In the $k^{\rm th}$ subfamily $(k=1,2,\ldots,R+1)$
we have vectors
labelled by four indices: $\ket{k;i;a,b}$. This vector has nonzero elements
only on the $i^{\rm th}$ level of the Parisi hierarchy, and there only
inside one column and row of blocks of size $p_{k-1}\times p_{k-1}$.
There are $n/p_{k-1}$
such columns, and the label $a=1,2,\ldots,n/p_{k-1}$ shows which is the one in
question. The vector components inside these blocks are all $1$'s,
except inside a distinguished column and row of blocks of size
$p_k\times p_k$ where they are
\footnote*{We take the opportunity
to correct a misprint in eq.~4 in [13] here.
The $B$ for $i=k-1$ was given as $B={1\over
2}\left(1-{p_{k-1}\over p_k}\right)$ there instead of the correct
expression in (25).}
$$\matrix{\displaystyle B={1\over
2}\left(2-{p_{k-1}\over p_k}\right),\hfill&\quad\quad{\rm if}\quad i=k-1\cr
\noalign{\vskip 8pt}
\displaystyle B=1-{p_{k-1}\over p_k},\hfill&\quad\quad{\rm
otherwise.}\hfill\cr}\eqno (25)$$
The distinguished columns of $p_k\times p_k$ blocks are labelled by the
last index, $b=1,2,\ldots,{p_{k-1}\over p_k}$.

With the choice (25) the vectors $\ket{k;i;a,b}$ are
orthogonal to all the previous subfamilies with $k-1,k-2,\ldots$ etc.~down to
the first family. Within the $k^{\rm th}$ subfamily vectors belonging to
different $i$'s and $a$'s are also orthogonal, while those with a
fixed $k,i,a$ and different $b$'s make a $({p_{k-1}\over p_k}-1)$-dimensional
hypertetrahedron. The number of linearly independent vectors for
a given $k$ and $i$ will then be ${n\over
p_{k-1}}\left({p_{k-1}\over p_k}-1\right)$, i.e.
$$\mu (k)=n\left({1\over p_k}-{1\over p_{k-1}}\right).\eqno (26)$$
The biorthogonal set associated with the tetrahedral groups
of vectors belonging to a given triplet $k,i,a$ is
$$\widetilde{\ket{k;i;a,b}}={p_k\over p_{k-1}^2}\,{1\over g^{(k)}_i}\,
\left(\ket{k;i;a,b}-\ket{k;i;a,{p_{k-1}\over p_k}}\right)\eqno (27)$$
where the weight $g^{(k)}_i$ is defined as
$$g^{(k)}_i=\cases {p_i-p_{i+1},\quad\quad&$i<k-1$\cr
\noalign{\vskip 3pt}
{1\over 4}(p_{k-1}-2p_k),\quad\quad&$i=k-1$\cr
\noalign{\vskip 3pt}
{1\over 2}(p_i-p_{i+1}),\quad\quad&$i>k-1$.\cr}\eqno (28)$$
For fixed $k,a,b$ the $R+1$ vectors $\ket{k;i;a,b}$,
$i=0,1,\ldots,R,$ form an invariant
subspace of any ultrametric matrix, so we will have eigenvectors
of the form
$$\sum_{i=0}^R\, f_k(i)\,\ket{k;i;a,b}\eqno (29)$$
with amplitudes $f_k(i)$ independent of $a$,$b$. The corresponding
eigenvalue equations will give $R+1$ possible values for $f_k(i)$, and
the eigenvalues $\lambda_m(k)$, $m=0,1,\ldots,R$,
will be $\mu(k)$-fold degenerate. Sometimes the second family is also
called the anomalous or A family.

So far
from the ${1\over 2}n(n-1)$-dimensional replica space we have split off the
$(R+1)$-dimensional invariant subspace of the first family, the
$(R+1)$-dimensional subspaces, $\mu(1)$ in number, of the $k=1$ subfamily,
etc., up to $k=R+1$, that is we have decomposed our linear space
into
$$\sum^{R+1}_{k=0}\,\mu(k)=n$$
$(R+1)$-dimensional invariant subspaces plus the vast
space, of dimension ${1\over 2}n(n-1)-n(R+1)$,
orthogonal to the first and second
families.
\medskip
\underbar{The third family}

The third family, often called the replicon or R family,
comprises everything remaining after
splitting off the first two families. It is a most remarkable
fact, and a direct consequence of the stringent conditions
ultrametricity imposes upon a matrix, that the third family can
be decomposed into invariant subspaces of
dimension $1$, i.e.~directly into eigenvectors.
This also means that the third
family eigenvalues that, for large $n$, represent the
overwhelming majority of all the eigenvalues can be obtained in
closed form, in terms of the matrix elements, for any ultrametric
matrix.

The third family eigenvectors were given in a concise
form in [13].
We provide a little more detail here which will
become important when we invert the matrix $M$.

The third family consists of several subfamilies labelled by three integers
$$\eqalign{r&=0,1,\ldots,R\cr k,l&=r+1,r+2,\ldots,R+1.\cr}$$
There will be several degenerate vectors in each subfamily. They
will be labelled by three, five, or seven more indices, as the
need arises. A common property of all third family vectors
is that they have nonzero components only inside one single
diagonal block of size $p_r\times p_r$, which also gives the significance of
the label $r$ above. The labels $k$ and $l$ specify further
structural details that are best displayed on a series of
figures again. In the following we shall exhibit only that
$p_r\times p_r$-sized block over which the vector components are not all
zero.

\underbar{The $r,\quad k=r+1,\quad l=r+1$ subfamily}

The structure of the nonvanishing block is shown in Fig.~5. These
vectors take three further labels to specify them $\ket{r;r+1,r+1;a,b,c}$.
The index
$a=1,2,\ldots,n/p_r$ shows which of the $n/p_r$ diagonal
$p_r\times p_r$ blocks has nonvanishing
elements. Inside this block all the components belonging to the
diagonal $p_{r+1}\times p_{r+1}$ blocks vanish again.
Of the off-diagonal $p_{r+1}\times p_{r+1}$ blocks
those belonging to two columns and rows are distinguished and a
further distinction is made between the blocks at the crossing
of a distinguished column and row and the rest. The indices
$b,c=1,2,\ldots,p_r/p_{r+1}$, $b\ne c$,
label the two distinguished columns. In all, we then have three
different vector components in this subfamily, as shown in the
figure.
\medskip
\vbox{\centerline{
\epsfxsize=5truecm\epsfbox{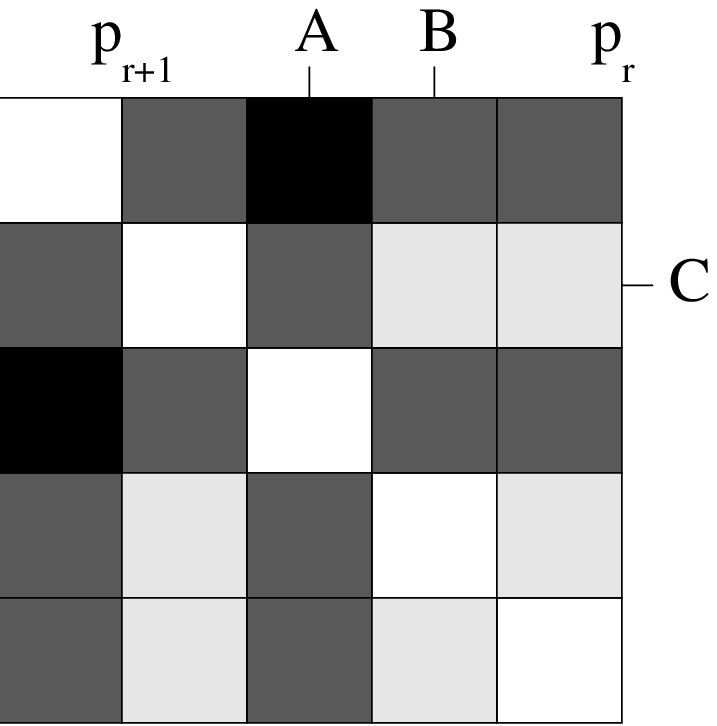}}
\smallskip
\noindent
Fig.~5.: The third family vector $\ket{r;r+1,r+1;a,b,c}$.
Identical shading means identical
components, blank means zero. $b=1,\,c=3$ in the figure.\par}
\bigskip\medskip
Orthogonality to the previous families requires that
$$\eqalign{\sum^{p_r/p_{r+1}}_{\eqalign{\scriptstyle b&
\scriptstyle =1\cr \noalign{\vskip -6pt} \scriptstyle b&\scriptstyle \ne c\cr}}
\,\ket{r;r+1,r+1;a,b,c}&=0\cr
\noalign{\vskip 5pt}
\sum^{p_r/p_{r+1}}_{\eqalign{\scriptstyle c&\scriptstyle =1\cr
\noalign{\vskip -6pt}
\scriptstyle  c&\scriptstyle \ne b\cr}}
\,\ket{r;r+1,r+1;a,b,c}&=0.\cr}\eqno(31)$$
It follows from (31) that the sum of vector components in
each row should vanish, giving us two equations for the three
numbers $A,B,C$, the third being determined by normalization. They
work out to be:
$$\eqalign{B&=-{p_{r+1}\over p_r-2p_{r+1}}\,A\cr
\noalign{\smallskip}
C&={2p_{r+1}^2\over (p_r-2p_{r+1})(p_r-3p_{r+1})}\,A\cr
\noalign{\smallskip}
A^2&={p_r-3p_{r+1}\over p_{r+1}^2(p_r-p_{r+1})}\,\,.\cr}\eqno (32)$$
With this we have determined the eigenvectors with $r,r+1,r+1$
completely. The corresponding eigenvalues will be written up in
the next section. For a given position of the $p_r\times p_r$ block, i.e.~
for a given $a$, the orthogonality conditions (31) leave
${1\over 2}{p_r\over p_{r+1}}\left({p_r\over p_{r+1}}-3\right)$
vectors linearly independent. This number, multiplied by $n\over
p_r$,
the number of choices for $a$, gives the multiplicity of this
class:
$$\mu(r;r+1,r+1)={1\over 2}\,n\,{p_r-3p_{r+1}\over p_{r+1}^2},\quad\quad
r=0,1,\ldots,R.\eqno (33)$$

\underbar{Eigenvectors with $r,\quad k>r+1,\quad l=r+1$}

An example is shown in Fig.~6.
\medskip
\vbox{\centerline{
\epsfxsize=6.5truecm\epsfbox{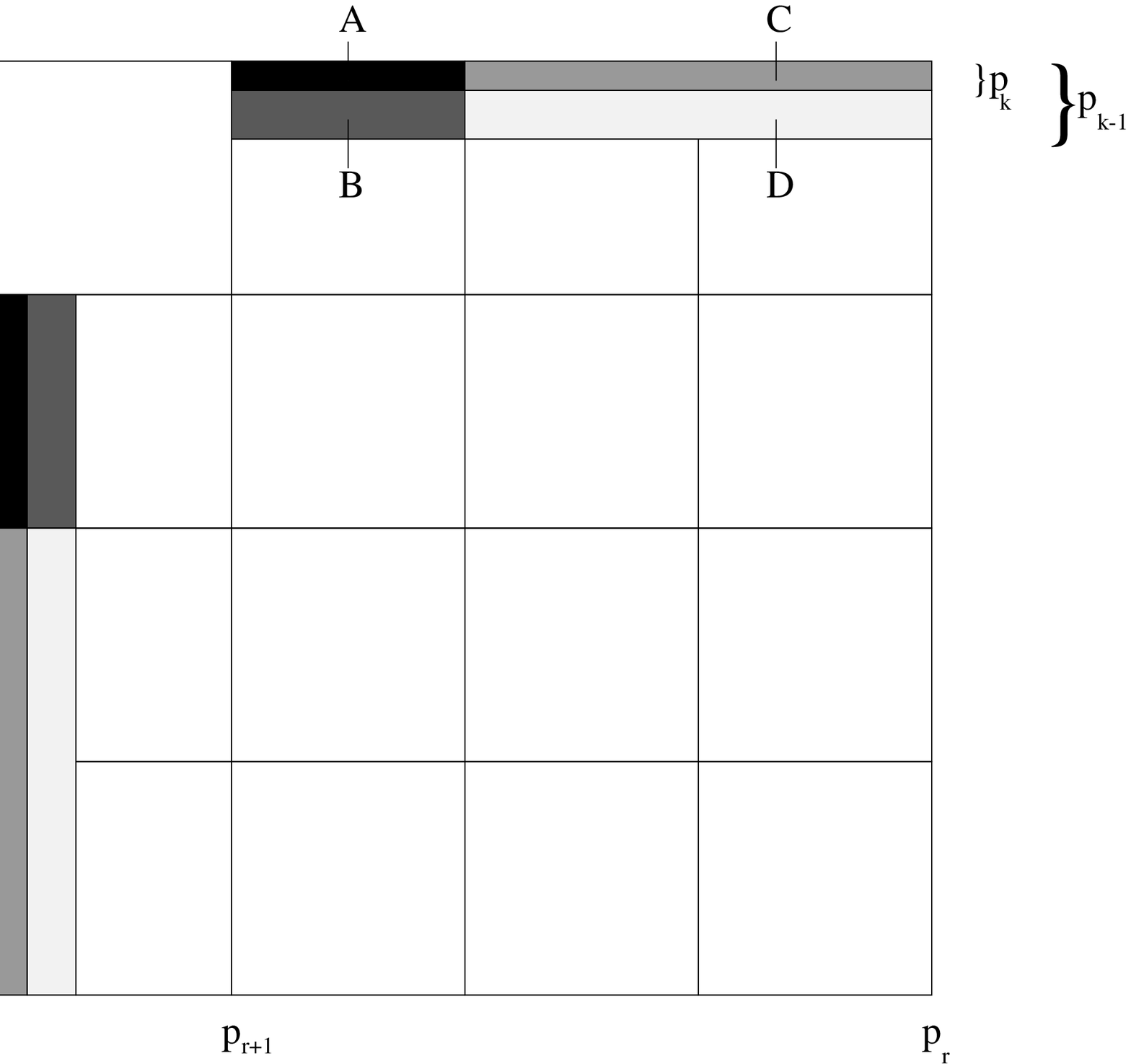}}
\smallskip
\noindent
Fig.~6.: An eigenvector of the class $\ket{r;k\ge r+2,l=r+1;a,b,c,d,e}$.\par}
\bigskip\medskip
Such a vector is constructed as follows. One chooses a diagonal
block of size $p_r\times p_r$, labelled by $a=1,2,\ldots,n/p_r$,
as before. Inside this block
one chooses two columns and rows of blocks of size
$p_{r+1}\times p_{r+1}$, say the
$b^{\rm th}$ and the $c^{\rm th}$, such that $c>b$ ($b=1$, $c=2$
in Fig.~6). Inside the blocks in the $b^{\rm th}$
column and row one now chooses a strip of blocks of size
$p_{k-1}\times p_{k-1}$,
say the $d^{\rm th}$, as shown in the figure. All the vector components
outside this strip will be zero. Inside the strip one chooses a
strip, the $e^{\rm th}$, of blocks of size $p_k\times p_k$. Finally the vector
components $A,B,C,D$ are arranged, as shown, according to whether they
belong to the strip of width $p_k$ or are outside, and also
whether they belong to the $c^{\rm th}$ blocks or not.

Orthogonality to
previous families again requires, as throughout the third
family, that the sum of components in each row vanish. This
immediately gives
$$\eqalign{B&=-{p_k\over p_{k-1}-p_k}\,A\cr
\noalign{\smallskip}
C&=-{p_{r+1}\over p_r-2p_{r+1}}\,A\cr
\noalign{\smallskip}
D&={p_kp_{r+1}\over (p_{k-1}-p_k)(p_r-2p_{r+1})}\,A\cr
\noalign{\smallskip}
A^2&={p_r-2p_{r+1}\over p_{r+1}(p_r-p_{r+1})}\left({1\over p_k}-
{1\over p_{k-1}}\right)\cr}\eqno (34)$$
for the components of a normalised vector of this class.

Considering the various choices for the parameters $a,b,c,d,e$ one finds
that the total number of linearly independent vectors of this
kind is
$$\mu(r;k,r+1)={1\over 2}\,n\,{p_r-2p_{r+1}\over p_{r+1}}\,
\left({1\over p_k}-{1\over p_{k-1}}\right),\quad\quad
k=r+2,r+3,\ldots,R+1.\eqno (35)$$

The subfamily $r;k=r+1,l>r+1$ is obtained by the same construction but
with $c<b$; this subfamily will be similar to the $r;k>r+1,l=r+1$ subfamily in
every respect, but
it will be orthogonal to it.

\underbar{Eigenvectors with $r,\quad k>r+1,\quad l>r+1$}

The construction of these vectors is shown in Fig.~7. It begins
again by choosing one diagonal block of size $p_r\times p_r$, labelled by
$a$. This can be done in $n/p_r$ different ways. Next, inside this
block one chooses two symmetrically positioned off diagonal
blocks of size $p_{r+1}\times p_{r+1}$.
This takes two indices: $b,c=1,2,\ldots,p_r/p_{r+1}$, and the number
of independent choices is ${1\over 2}{p_r\over
p_{r+1}}\left({p_r\over p_{r+1}}-1\right)$, because of the symmetry of the
matrix representing the eigenvector. Now the off diagonal
$p_{r+1}\times p_{r+1}$
block above the diagonal is cut into rectangles of horizontal
size $p_{l-1}$ and vertical size $p_{k-1}$ and the one below the diagonal is
cut similarly with the horizontal and vertical dimensions
exchanged.
\medskip
\vbox{\centerline{
\epsfxsize=\hsize\epsfbox{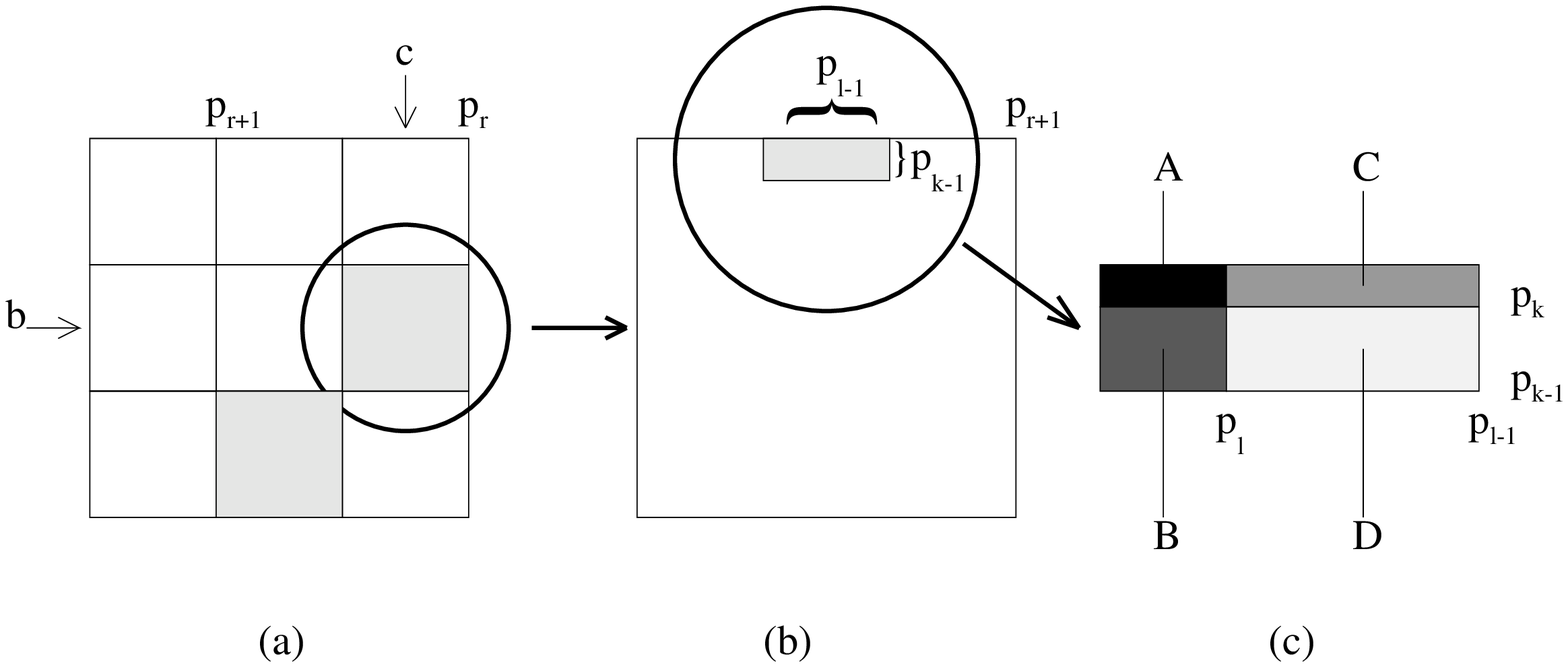}}
\smallskip
\noindent
Fig.~7.: The construction of an eigenvector with $r,k>r+1,l>r+1$. The
structure of the vector is shown in three stages of finer and
finer resolution. The block with indices $c,b$ in (a) is obtained
by reflection to the diagonal from block $b,c$.\par}
\bigskip\medskip
One of these rectangles is chosen, which again takes two labels:
$d=1,2,\ldots,p_{r+1}/p_{l-1}$, and $e=1,2,\ldots,p_{r+1}/p_{k-1}$,
and can be done in $p_{r+1}^2/(p_{l-1}p_{k-1})$ ways. We have nonzero
components only inside these rectangles. Their structure is
shown in Fig.~7(c), and can evidently be characterised by two
further indices $f=1,2,\ldots,p_{l-1}/p_l$, and $g=1,2,\ldots,p_{k-1}/p_k$.
These vectors are orthogonal to
each other in all the indices except for $f$ and $g$. For fixed $f$ the
set with different $g$'s forms the usual tetrahedron again, and
the same is true for fixed $g$ in the $f$'s, so we are left with
$({p_{k-1}\over p_k}-1)\times ({p_{l-1}\over p_l}-1)$ independent choices
for $f$ and $g$. All these taken together
give a multiplicity
$$\mu(r;k,l)={1\over 2}n\,(p_r-p_{r+1})\,\left({1\over
p_k}-{1\over p_{k-1}}\right)\,\left({1\over
p_l}-{1\over p_{l-1}}\right),\quad\quad
k,l=r+2,r+3,\ldots,R+1,\eqno (36)$$
while the usual orthogonality conditions (the sum of vector
components in each row and each column vanishes) give us the
following values for the components of a normalised vector of
the $r,k>r+1,l>r+1$ type:
$$\eqalign{B&=-{p_k\over p_{k-1}-p_k}\,A\cr
C&=-{p_{l}\over p_{l-1}-p_{l}}\,A\cr
D&={p_kp_{l}\over (p_{k-1}-p_k)(p_{l-1}-p_{l})}\,A\cr
A^2&=\left({1\over p_k}-
{1\over p_{k-1}}\right)\,\left({1\over p_l}-
{1\over p_{l-1}}\right).\cr}\eqno (37)$$

With this we have given a full description of the third
family eigenvectors. To show that these forms are indeed
reproduced under the action of an ultrametric matrix takes, of
course, a lot of algebra. It is impossible for us to go into
details on this, but we think a little experimentation in a
simple special case like $R=1$ will convince the reader that
the proof is quite straightforward though certainly not very
short.  As we have seen, there is a high degree of degeneracy
within each subfamily $(r;k,l)$. These subfamilies are all orthogonal
to each other (and also to the first and second families, of
course), and some of the degenerate vectors within a given $(r;k,l)$
subfamily are also orthogonal, but some form the by now familiar
tetrahedral sets in more than one index. It would not be difficult to
orthogonalise these vectors, or, alternatively, to construct
biorthogonal sets to them. We refrain from doing both: the loss in
symmetry would be considerable and the gain virtually nothing.
The only occasion when we might need a properly orthonormalised
set is when later we construct the "replicon" components of the
inverse of $M$ from the spectral resolution. It will turn out, however,
that the orthogonalisation can be circumvented even there and
the ultrametric symmetries of $M$ (and of its inverse) will
allow us to deduce the full contribution of the whole $(r;k,l)$
subfamily to the inverse from the knowledge of a single vector
belonging to that subfamily. This vector will be called the
{\bf representative vector} of the subfamily.

We can choose any of
the $\mu(r;k,l)$ degenerate vectors to be the representative vector.
Suppose we have made our choice. Some of the vectors in the $(r;k,l)$
subfamily will be orthogonal to the selected vector from the beginning.
These will, as a rule, have zero components where the selected
vector has nonzero ones, in particular, they will have zeros
where the components we called $A$'s in the description of the
third family vectors, i.e.~those in the darkest shaded areas in
the figures, are to be found in the selected vector. Now
consider the vectors which are not orthogonal to the selected
one. In order to orthogonalise them to the representative vector
we form some linear combinations. Our key observation is now
that, due to the defining properties of the third family vectors
(namely, that the sums of vector components in each row are zero),
any linear combination that is orthogonal to the representative
vector will have zero components where the representative vector
has its components $A$. The proof of this will also be left as
an exercise to the reader.

The last issue to be settled in this
section is the total multiplicity of our basis vectors.
In the first two families we found $n\,(R+1)$ independent basis
vectors. In the subfamilies $(r;k,l)$, for fixed $r$, there are a
total of
$$\sum_{k,l=r+1}^{R+1}\mu(r;k,l)={1\over 2}n(p_r-p_{r+1}-2)\eqno
(38)$$
independent vectors. This summed over $r$ gives the total
number of basis vectors in the third family:
$$\sum_{r=0}^R{1\over 2}n(p_r-p_{r+1}-2)={1\over
2}n(n-2R-3).\eqno (39)$$
Added to $n\,(R+1)$ this gives ${1\over 2}n(n-1)$, the dimension of replica
space. Our set of basis vectors is therefore complete.
\medskip\bigskip

\noindent{\bf 4. The block diagonal form of ultrametric matrices}
\bigskip
Having constructed a complete set of basis vectors we can build
a matrix $S$ with columns made of these vectors and transform
$M$ to this new basis by
$$\tilde M=S^{-1}MS.\eqno (40)$$
$S$ is not an orthogonal transformation (because the basis
vectors are not all orthogonal), so $\tilde M$ will not be symmetric.
The rows of the inverse $S^{-1}$ in the first family sector will be
made of the bra vectors corresponding to the first-family-like
basis vectors $\ket{0;i}$, in the second family sector they will be the
biorthogonal vectors given in (27). In the third family sector
we do not really need to construct the matrix $S^{-1}$ at all, since
there the basis vectors are the eigenvectors themselves. Since
the various families and subfamilies were constructed in such a
way that they are invariant subspaces of $M$, the transformed
matrix $\tilde M$ will have a block diagonal form: along the diagonal
we will have a string of $n$ $(R+1)\times (R+1)$ matrices,
the first corresponding
to the first family, the next $\mu(1)$ identical matrices
corresponding to the $k=1$ subfamily in the second family, etc.~
through $\mu(k)$ identical blocks for the $k^{\rm th}$ subfamily up to $k=R+1$.
This string of matrices will be followed by the string of the
third family eigenvalues coming in groups of $\mu(r;k,l)$ identical
numbers corresponding to the subfamilies $(r;k,l)$.

The third family
or replicon eigenvalues are obtained as a byproduct of checking
that the third family vectors given in the previous section are
eigenvectors indeed. In the $(r;k,l)$ subfamily one obtains the closed
expression
$$\eqalign{\lambda(r;k,l)&=\sum^{R+1}_{s=k}p_s\sum^{R+1}_{t=l}p_t
(M^{r,r}_{t,s}-M^{r,r}_{t-1,s}-M^{r,r}_{t,s-1}+M^{r,r}_{t-1,s-1})\cr
r&=0,1,\ldots,R\cr k,l&=r+1,r+2,\ldots,R+1,\cr}\eqno (41)$$
giving the replicon eigenvalues directly in terms of the
matrix elements and of the $p$'s characterising the structure of
$M$. Eq.~(41) has been written up already in [13].

The $(R+1)\times (R+1)$
diagonal blocks of $\tilde M$ will be labelled by $k=0,1,\ldots,R+1$
as $M^{(k)}$, $k=0$
corresponding to the first family, $k>0$ to the subfamilies in the
second family. The matrix elements of $M^{(0)}$ can be obtained by
sandwiching $M$ between two first family vectors
$$M^{(0)}_{r,s}=\bra{0;r}M\ket{0;s},\eqno (42)$$
those of $M^{(k)}$ by sandwiching $M$ between a second family
vector and one from the biorthogonal set given in (27):
$$M^{(k)}_{r,s}=\widetilde{\bra{k;r;a,b}}M\ket{k;s;a,b}\eqno (43)$$
and can really be obtained again as byproducts
when verifying the invariance of
the various subfamilies under the action of the matrix
$M$.

We can therefore simply state the result:
$$M^{(k)}_{r,s}=\Lambda(k,r)\Kr{r}{s}+g_s^{(k)}
{\left(\Delta_r^{(k)}+{1\over 2}p_{k-1}\Kr{r}{k-1}\right)
\left(\Delta_s^{(k)}+{1\over 2}p_{k-1}\Kr{s}{k-1}\right)\over
\d_r\d_s}K_k(r,s).\eqno (44)$$
Of the symbols appearing here $\d_r$ and $g_s^{(k)}$ have already been
defined in (16) and (28), respectively. $\Kr{r}{s}$ is the Kronecker
symbol, while $\Delta_r^{(k)}$ is
$$\Delta_r^{(k)}=\cases{{1\over 2}\d_r,&$r<k-1$\cr
\noalign{\smallskip}
{1\over2}(\d_{k-1}-p_k),\quad\quad&$r=k-1$\cr
\noalign{\smallskip}
\d_r,&$r>k-1$.\cr}\eqno(45)$$
The diagonal part $\Lambda(k,r)$ is related to the third family
eigenvalues:
$$\Lambda(k,r)=\cases{\lambda(r;r+1,k),&$k>r+1$\cr
\lambda(r;r+1,r+1),&$k\le r+1$.\cr}\eqno(46)$$

Now we define the {\bf kernel} $K_k(r,s)$ appearing in (44):
$$\eqalign{{1\over4}K_k(r,s)&={1\over4}\sum^r_{j=k}p_j(\M{r}{s}{j}{j}-
\M{r}{s}{j-1}{j-1})+{1\over2}\sum^s_{j=r+1}p_j(\M{r}{s}{r}{j}-
\M{r}{s}{r}{j-1})\cr\noalign{\smallskip}
&+\sum^{R+1}_{j=s+1}p_j(\M{r}{s}{r}{j}-
\M{r}{s}{r}{j-1}),\qquad\qquad k-1\le r\le s,\cr}\eqno(47)$$
$$\eqalign{{1\over4}K_k(s,r)&={1\over4}\sum^r_{j=k}p_j(\M{s}{r}{j}{j}-
\M{s}{r}{j-1}{j-1})+{1\over2}\sum^s_{j=r+1}p_j(\M{s}{r}{j}{j}-
\M{s}{r}{j-1}{j-1})\cr\noalign{\smallskip}
&+\sum^{R+1}_{j=s+1}p_j(\M{s}{r}{s}{j}-
\M{s}{r}{s}{j-1}),\qquad\qquad k-1\le r\le s,\cr}\eqno(48)$$
$${1\over4}K_k(r,s)={1\over2}\sum^s_{j=k}p_j(\M{r}{s}{r}{j}-
\M{r}{s}{r}{j-1})+\sum^{R+1}_{j=s+1}p_j(\M{r}{s}{r}{j}-
\M{r}{s}{r}{j-1}),\quad r\le k-1\le s,\eqno(49)$$
$${1\over4}K_k(s,r)={1\over2}\sum^s_{j=k}p_j(\M{s}{r}{j}{j}-
\M{s}{r}{j-1}{j-1})+\sum^{R+1}_{j=s+1}p_j(\M{s}{r}{s}{j}-
\M{s}{r}{s}{j-1}),\quad r\le k-1\le s,\eqno(50)$$
$${1\over4}K_k(r,s)=\sum^{R+1}_{j=k}p_j(\M{r}{s}{r}{j}-
\M{r}{s}{r}{j-1}),\quad r\le s\le k-1,\eqno(51)$$
$${1\over4}K_k(s,r)=\sum^{R+1}_{j=k}p_j(\M{s}{r}{s}{j}-
\M{s}{r}{s}{j-1}),\quad r\le s\le k-1.\eqno(52)$$
(Possible empty sums here and in the following are understood to
be zero. For $k=0$, terms with $\M rs{-1}{-1}$ may occur
in the above formulae. They are, by definition, zero too.)

With this the matrix elements $M^{(k)}_{r,s}$ in the new representation
have
been expressed in terms of the matrix elements in the original
representation. The problem of finding the eigenvalues of $M$ in
the first two families has thus been broken down into the
problem of finding the spectrum of each of the $M^{(k)}$'s. As we have
already noted, due to the nonorthogonality of the transformation
$S$, the $M^{(k)}$'s are not symmetric. Using the symmetries between the
various components of $M$ as described in Sec.~2, one can show,
however, that, although we have given the expressions for both $K_k(r,s)$
and $K_k(s,r)$ in the various cases for completeness,
the kernel $K_k$
is, in fact, symmetric. Therefore, the asymmetry of $M^{(k)}$ is
carried solely by the factor $g^{(s)}_k$ in eq.~(44), and we can, if we
wish, reduce the eigenvalue problem of $M^{(k)}$ to that of the
manifestly symmetric matrix
$${g^{(k)}_r}^{1\over2}\,{M^{(k)}_{r,s}
\over g^{(k)}_s}\,{g^{(k)}_s}^{1\over2}.$$

Let us now spell out the
eigenvalue equation of $M^{(k)}$:
$$\sum_{s=0}^RM^{(k)}_{r,s}f_k(s)=\lambda(k)f_k(r)$$
reads in the two cases $r<k-1$ and
$r\ge k-1$, respectively, as
$$\eqalign{\Lambda(k,r)f_k(r)+{1\over4}\sum^R_{\eqalign{\scriptstyle
s&\scriptstyle =0\cr
\noalign{\vskip -6pt}
\scriptstyle s\ne&\scriptstyle  k-1\cr}}K_k(r,s)f_k(s)\d_s&+
{1\over 8}(\d_{k-1}-p_k)
K_k(r,k-1)f_k(k-1)\cr
&=\lambda(k)f_k(r),\quad\quad r<k-1\cr
\noalign{\smallskip}
\Lambda(k,r)f_k(r)+{1\over2}\sum^R_{\eqalign{\scriptstyle s&\scriptstyle
=0\cr
\noalign{\vskip -6pt}
\scriptstyle s \ne&\scriptstyle k-1\cr}}K_k(r,s)f_k(s)\d_s&+
{1\over 4}(\d_{k-1}-p_k)
K_k(r,k-1)f_k(k-1)\cr
&=\lambda(k)f_k(r),\quad\quad r\ge k-1.\cr}\eqno(53)$$
This is a set of $R+1$ homogeneous linear equations for any
given $k=0,1,\ldots,R+1$. (In the first family, i.e.~for $k=0$,
only the equation
with $r>k-1$ applies, with the third term on the l.h.s.~discarded.)
The solutions for the $f_k$'s give the amplitudes mentioned in
the preceding section, while the $\lambda(k)$'s are the
corresponding eigenvalues.

It is evident from the definition that the space of ultrametric
matrices (belonging to the same series $p_0,p_1,\ldots,p_R$) is closed under
addition. That it is also closed under multiplication is
easiest to see from the existence of the common similarity
transformation $S$ that brings any two such matrices to
blockdiagonal form simultaneously.

Suppose we are given two
ultrametric matrices $M$ and $M'$, with the associated kernels $
K_k$
and $K_k'$ and replicon eigenvalues $\lambda^R$ and $\lambda^{R'}
$
from which we have
the corresponding $\Lambda$'s as given in (46). Then in the block
diagonal representation of the product $MM'$ we will find for the
diagonal blocks in the LA sector:
$$\eqalign{\sum_{t=0}^RM^{(k)}_{r,t}{M^{(k)}_{t,s}}'&=g_s^{(k)}
{\left(\Delta_r^{(k)}+{1\over 2}p_{k-1}\Kr{r}{k-1}\right)
\left(\Delta_s^{(k)}+{1\over 2}p_{k-1}\Kr{s}{k-1}\right)\over
\d_r\d_s}\times\cr&
\left\{{1\over2}\sum^R_{t=0}\Delta^{(k)}_tK_k(r,t)
K_k'(t,s)+\Lambda(k,r)K_k'(r,s)
+\Lambda'(k,s)K_k(r,s)\right\}\cr
\noalign{\vskip 10pt}&
+\Kr{r}{s}\Lambda(k,r)\Lambda'(k,r)\cr}\eqno(54)$$
where we
have used that by (16), (28), (45)
$$g_t^{(k)}
{\left(\Delta_t^{(k)}+{1\over 2}p_{k-1}\Kr{t}{k-1}\right)^2\over
\d_t^2}\equiv {1\over2}\Delta^{(k)}_t,\eqno(55)$$
whereas in the
$R$ sector we will evidently find the product of the replicon
eigenvalues:
$$\lambda(r;k,l)\lambda'(r;k,l).\eqno(56)$$
In particular, if $M'$ is the inverse of
$M$, we have $\sum\limits^R_{t=0}M^{(k)}_{r,t}M'^{(k)}_{t,s}=\Kr{r}{s}$
and $\Lambda(k,r)\Lambda'(k,r)=1$, so from (54) we get
$${1\over2}\sum^R_{t=0}\Delta^{(k)}_tK_k(r,t)
K_k'(t,s)+\Lambda(k,r)K_k'(r,s)
+\Lambda'(k,s)K_k(r,s)=0.\eqno(57)$$
For a
given $M$, i.e.~for a given $K_k$ and $\Lambda(k,r)$,
this is an equation for
the kernel $K_k'$ of the inverse matrix. In the general case (57)
is still a matrix equation which is typically difficult to solve,
though (57) certainly has the merit of reducing the problem of
the inversion of a ${1\over2}n(n-1)\times {1\over2}n(n-1)$
dimensional matrix to that of inverting
$R+2$ much smaller matrices (corresponding $k=0,1,\ldots,R+1$ in (57)) of size
$(R+1)\times (R+1)$. In some important special cases, however, further progress
can be made and both the solution of the eigenvalue equations
(53) and the inversion of the matrix $M$ can be carried
through to the end [12].

Coming back to eq.~(57) let us assume now
that we have somehow succeeded in solving it for $K'_k$. With this the
inversion of $M$ is, however, not yet completed, because
normally we need the inverse in the original Cartesian
coordinate system, i.e.~we need the components of $M'$ given the
kernel $K'_k$. This means we have to invert the formulae (47)-(52),
or, to put it even more simply, we have to turn (40) around like
$$M=S\tilde MS^{-1}.\eqno(58)$$

In the LA sector this is a standard operation: we
have all the basis vectors and their biorthogonal counterparts
so that we explicitly know the corresponding blocks in $S$ and
$S^{-1}$. This is not the case in the R sector where we have
neither orthogonalised nor biorthogonalised our basis vectors.
In the direct transformation, from $M$ to $\tilde M$, this did not
cause a problem, because the third family basis vectors being
eigenvectors we knew in advance that the corresponding "blocks"
of $\tilde M$ would be the eigenvalues themselves. In the inverse
transformation, from $\tilde M$ to $M$, however, we would definitely need
the missing blocks in $S^{-1}$ in order to determine the contribution
of the replicon family to the various components of $M$. It is
at this point that the concept of the representative vector
introduced in the previous section becomes important. We do not
think we should dwell upon how the blocks of the three matrices
in (58) have to be multiplied in the sector where they are known.
We have to explain, however, how the replicon
contribution to (58) can be obtained from the representative
vectors without actually knowing the corresponding block in $S^{-1}$.
The Appendix is devoted to this problem.

In what follows we will
state our results for the nine different types of
matrix elements of $M$ discussed in Sec.~2, in terms of the
matrix elements of the blockdiagonal form $\tilde M$. In each case we
shall give the result in two different forms: first as a sum of
two terms, one coming from the LA sector, the other from the
replicon, and second, in a form where some most remarkable
cancellations between these two have been effected. In the
discrete case, where $n$, $R$, and all the $p_k$'s are integers, these
cancellations may seem coincidental. We note, however, that in
the continuous limit they acquire a
fundamental importance [13].

In order to display these cancellations we
partition the third family multiplicities as follows:
$$\mu(r;k,l)\equiv\mu_{\rm reg}(r;k,l)+\mu_{\rm sing}(r;k,l)\eqno(59)$$
where
$$\eqalignno{\mu_{\rm reg}(r;r+1,r+1)&={1\over 2}n{p_r\over
p_{r+1}}\left({1\over p_{r+1}}-{1\over p_r}\right),&(60)\cr
\mu_{\rm sing}(r;r+1,r+1)&=-\sum^{r+1}_{k=0}\mu(k),&(61)\cr}$$
$$\eqalign{\mu_{\rm reg}(r;r+1,k)=\mu_{\rm reg}(r;k,r+1)&=
{1\over 2}n{p_r-p_{r+1}\over p_{r+1}}\left({1\over p_k}-{1\over
p_{k-1}}\right),\cr
\noalign{\medskip}
&\qquad\qquad\qquad k>r+1,\cr}\eqno(62)$$
$$\mu_{\rm sing}(r;r+1,k)=\mu_{\rm sing}(r;k,r+1)=-{1\over 2}\mu(k),
\qquad k>r+1\eqno(63)$$
and $\mu_{\rm reg}(r;k,l)$ for $k,l>r+1$ is the full
$\mu (r;k,l)$ itself, as given in (36), so
$$\mu_{\rm sing}(r;k,l)=0,\qquad\qquad k,l>r+1.\eqno(64)$$

In (61) and (63) $\mu(k)$ is the second family multiplicity.
In the discrete case the subscripts "regular" and "singular"
have no particular significance; in the continuous limit,
however, $\mu_{\rm sing}$ will be associated with terms that become
meaningless but disappear from the theory due to the
cancellations mentioned above.

We now list the results:
$$\M rr{R+1}{R+1}=\sum_{k=0}^{R+1}{2\mu (k)\over n\d_r}M^{(k)}_{r,r}
+\sum_{k=r+1}^{R+1}\sum_{l=r+1}^{R+1}{2\mu (r;k,l)\over n\d_r}
\lambda (r;k,l)\eqno (65a)$$
where the first term is the LA, the second the R
contribution as announced. Substituting (44) for $M^{(k)}$ and
splitting $\mu$ as $\mu=\mu_{\rm reg}+\mu_{\rm sing}$ we see
that the $\Lambda(k,r)$ term coming from the LA cancels
the $\mu_{\rm sing}$
contributions from the R family exactly.  So we have the alternative
form:
$$\M rr{R+1}{R+1}=\sum_{k=0}^{R+1}{\mu (k)\over n\d_r}\Dk r\F rr
+\sum_{k=r+1}^{R+1}\sum_{l=r+1}^{R+1}{2\mureg rkl\over n\d_r}
\lambda (r;k,l).\eqno (65b)$$
Similarly:
$$\eqalignno{\M rr{R+1}{t}&=\sum_{k=0}^{R+1}{2\mu (k)\over n\d_r}
{\Dk t\over \d_t}M^{(k)}_{r,r}&\cr
&+\sum_{k=r+1}^{R+1}\sum_{l=r+1}^{R+1}{2\mu (r;k,l)\over n\d_r}
\lambda (r;k,l)\left(\Dd kt+\Dd lt-1\right)&(66a)\cr
&=\sum_{k=0}^{R+1}{\mu (k)\over n}{\Dk r\Dk t\over\d_r\d_t}\F rr&\cr
&+\sum_{k=r+1}^{R+1}\sum_{l=r+1}^{R+1}{2\mureg rkl\over n\d_r}
\lambda (r;k,l)\left(\Dd kt+\Dd lt-1\right),\qquad t>r.&(66b)\cr}$$

In the special case $t=r$ of the above, things work out slightly differently:
$$\eqalignno{\M rr{R+1}{r}&=\cr
\noalign{\smallskip}
&=\sum_{k=0}^{R+1}{2\mu (k)\over n\d_r}
{\Dk r-{1\over 4}p_{k-1}\Kr r{k-1}\over \d_r-{1\over 2}p_{k-1}\Kr r{k-1}}
M^{(k)}_{r,r}\cr
&-{p_{r+1}\over p_r-2p_{r+1}}\Bigl\{{2\mu (r;r+1,r+1)\over n\d_r}
\lambda (r;r+1,r+1)\cr
&+\sum_{k=r+2}^{R+1}{2\mu (r;r+1,k)\over n\d_r}
\lambda (r;r+1,k)\Bigr\}&(67a)\cr
&=\sum_{k=0}^{R+1}{\mu (k)\over n\d_r}
{\Dk r-{1\over 4}p_{k-1}\Kr r{k-1}\over \d_r-{1\over 2}p_{k-1}\Kr r{k-1}}
\Dk r\F rr,&(67b)\cr}$$
so that, as we see, the $\Lambda$ term coming from the LA now cancels
the whole replicon contribution, not just the one with $\mu_{\rm
sing}$.

The next item to be considered is a component of $M$ with
different upper indices, so that there is no replicon contribution to
it:
$$\eqalignno{\M rs{R+1}{r}&=\sum_{k=0}^{R+1}{\mu (k)\over n}
{
\left(\d_r-{1\over 2}p_{k-1}\Kr r{k-1}\right)
\left(\d_s-{1\over 2}p_{k-1}\Kr s{k-1}\right)\over \d_r\d_s}
{M^{(k)}_{r,s}\over g^{(k)}_s}&\cr
&=\sum_{k=0}^{R+1}{\mu (k)\over n}{\Dk r\Dk s\over\d_r\d_s}\F rs,\qquad
s\ne r.&(68)\cr}$$

Now we turn to the matrix elements of the third
kind:
$$\eqalignno{\M rstt&=\sum^{R+1}_{k=0}{2\mu(k)\over n\d_s}
\left(2\Dd kt-1\right)M^{(k)}_{r,s}&\cr
&=\sum^{R+1}_{k=0}{\mu(k)\over n}\F rs
\left(2\Dd kt-1\right),\qquad t<r,s.&(69)\cr}$$
Note that this holds even for $r=s$, because
$$\sum^{R+1}_{k=0}\mu(k)
\left(2\Dd kt-1\right)=0,$$
therefore the diagonal part $\Lambda$ in $M^{(k)}$ never contributes. Neither
does the replicon, because the upper indices are larger than the lower ones
in here.

The next four items are off-diagonal
in the upper indices, so they do not receive contributions from the
replicon family.
$$\eqalignno{\M rs{t}{t}&=\sum_{k=0}^{R+1}{\mu (k)\over n}
{\left(\d_s-{1\over 2}p_{k-1}\Kr s{k-1}\right)
\over \d_s}
\left(2\Dd kt-1\right){M^{(k)}_{r,s}\over g^{(k)}_s}&\cr
&=\sum_{k=0}^{R+1}{\mu (k)\over n}{\Dk s\over \d_s}
\left(2\Dd kt-1\right)\F rs,\qquad
s<t<r,&(70)\cr}$$
$$\eqalignno{\M rs{r}{t}&=\sum_{k=0}^{R+1}{\mu (k)\over n}
{
\left(\d_r-{1\over 2}p_{k-1}\Kr r{k-1}\right)
\left(\d_s-{1\over 2}p_{k-1}\Kr s{k-1}\right)\over \d_r\d_s}
\left(2\Dd kt-1\right){M^{(k)}_{r,s}\over g^{(k)}_s}&\cr
&=\sum_{k=0}^{R+1}{\mu (k)\over n}{\Dk r\Dk s\over\d_r\d_s}
\left(2\Dd kt-1\right)\F rs,\qquad\qquad
r<s<t,&(71)\cr}$$
$$\eqalign{\M rs{r}{r}&=\sum_{k=0}^{R+1}{\mu (k)\over n}
\left(2\Dd kr-1\right){M^{(k)}_{r,s}\over g^{(k)}_s}\cr
&=\sum_{k=0}^{R+1}{\mu (k)\over n}
\left(2\Dd kr-1\right)\F rs,\qquad\qquad\qquad\hfill
r<s,\cr}\eqno(72)$$
$$\eqalign{\M rs{r}{s}&=\sum_{k=0}^{R+1}{\mu (k)\over n}
{\d_r-{1\over 2}p_{k-1}\Kr r{k-1}
\over \d_r}
\left(2\Dd ks-1\right){M^{(k)}_{r,s}\over g^{(k)}_s}\cr
&=\sum_{k=0}^{R+1}{\mu (k)\over n}{\Dk r\over\d_r}
\left(2\Dd ks-1\right)\F rs,\quad\quad\qquad
r<s.\cr}\eqno(73)$$

In the next two cases again a complete cancellation takes place
between the $\Lambda$ term in the LA and the replicon:
$$\eqalignno{\M rr{r}{r}&=\sum_{k=0}^{R+1}{2\mu (k)\over n\d_r}
{\Dk r+{1\over 2}p_{k-1}\Kr r{k-1}\over \d_r-{1\over 2}p_{k-1}\Kr r{k-1}}
\left(2\Dd kr-1\right)M^{(k)}_{r,r}&\cr
&+{2\mu (r;r+1,r+1)\over n\d_r}
{2p_{r+1}^2\over (p_r-2p_{r+1})(p_r-3p_{r+1})}\lambda (r;r+1,r+1)&(74a)\cr
&=\sum_{k=0}^{R+1}{\mu (k)\over n}\left(2\Dd kr-1\right)K_k(r,r),
&(74b)\cr}$$
$$\eqalignno{\M rr{r}{s}&=\sum_{k=0}^{R+1}{2\mu (k)\over n\d_r}\,
{\Dk r-{1\over 4}p_{k-1}\Kr r{k-1}\over \d_r-{1\over 2}p_{k-1}\Kr r{k-1}}
\left(\Dd kr+\Dd ks-1\right){\d_r\over \Dk r}
M^{(k)}_{r,r}&\cr
&-{p_{r+1}\over p_r-2p_{r+1}}\Bigl\{{2\mu (r;r+1,r+1)\over n\d_r}
\lambda (r;r+1,r+1)&\cr
&+\sum_{k=r+2}^{R+1}{2\mu (r;r+1,k)\over n\d_r}
\lambda (r;r+1,k)\,
{\Dk r\d_s+\Dk s\d_r-\d_r\d_s\over \Dk r\d_s}\Bigr\}&(75a)\cr
&=\sum_{k=0}^{R+1}{\mu (k)\over n}\,
{\Dk r-{1\over 4}p_{k-1}\Kr r{k-1}\over \d_r-{1\over 2}p_{k-1}\Kr r{k-1}}
\left(\Dd kr+\Dd ks-1\right)\F rr,\qquad s>r.&(75b)\cr}$$

Finally, in the last type of component the $\Lambda$ term from the LA
cancels the $\mu_{\rm sing}$ term in the R:
$$\eqalignno{\M rrst&=\sum_{k=0}^{R+1}{2\mu (k)\over n\d_r}
\left(\Dd ks+\Dd kt-1\right)M^{(k)}_{r,r}&\cr
&+\sum^{R+1}_{k=r+1}\sum^{R+1}_{l=r+1}{2\mu (r;k,l)\over n\d_r}
\left(2\Dd ks -1\right)\left(2\Dd lt -1\right)
\lambda(r;k,l)&(76a)\cr
&=\sum ^{R+1}_{k=0}{\mu (k)\over n\d_r}\Dk r\left(\Dd ks+\Dd kt-1\right)
\F rr &\cr
&+\sum^{R+1}_{k=r+1}\sum^{R+1}_{l=r+1}{2\mu_{\rm reg} (r;k,l)\over n\d_r}
\left(2\Dd ks -1\right)\left(2\Dd lt -1\right)
\lambda(r;k,l)\quad s,t>r.&(76b)\cr}
$$

As we have mentioned, in the equations from (65) to (76) the
formulae denoted (a) give the true partition of the
contributions between the LA and R families.  If anyone tried
to reproduce these results, they would inevitably get them in this
form, and we give them here partly as signposts. In most
applications the origin of the terms is completely immaterial,
however, so that when using these formulae, one will clearly apply
the (b) forms, where the cancellations have been performed. The
names one attaches to these terms are also largely a matter of
convention: in the papers [14] and [15] where analogous formulae
were derived for the propagators two of us used the name LA for
the first terms and the name R for the second terms {\bf in the (b)
forms}.

We also see that there is nothing mysterious about the
cancellations: the diagonal matrix elements of $M^{(k)}$ contain the
replicon eigenvalue and this piece partially or completely
cancels the contribution from the R family. In [13] two of us,
discussing the importance of these cancellations in the context
of the propagators, made the remark that a certain
asymptotic relation between the second family and third family
eigenvalues was a necessary condition for the cancellations to
work. Although the asymptotic relation between the eigenvalues
was certainly valid in the specific example discussed there, and
may be valid in more general situations also, we can clearly see
that it has nothing to do with the cancellations: these are a
purely "kinematic" effect, depending solely on the ultrametric
geometry and on no further details of the theory.

To conclude,
we make an additional remark. Before presenting formulae
(65)-(76) we gave a sketchy indication (with some details to be
added in the Appendix) as to how they can be obtained, which is, of
course, not necessarily the most economic way that they can be
verified once known. Eqs.~(65)-(76) are the inversion of
(47)-(52) and of (41). The simplest way to check them is by
direct substitution.

Having established the inverse relations
between the matrix elements and the kernel we can now summarise
the steps one has to follow in order to invert an ultrametric
matrix. First one has to determine the kernel and the replicon
eigenvalues of the matrix by (41), (47)-(52). To get the
replicon eigenvalues of the inverse matrix is trivial, they are
the reciprocal of the original replicon eigenvalues. To obtain
the new kernel requires the solution of (57). This is the hard
core that remains to be cracked after the layer controlled by
ultrametricity has been peeled off. To find the new kernel
requires the concrete knowledge of the matrix elements and as
such it is outside the scope of the present paper. Assuming the
new kernel has been found one finally obtains the elements of
the inverse matrix via eqs.~(65)-(76).
\bigskip\medskip
{\bf Acknowledgement.} We are grateful to M.~M\'ezard for
persistently urging us to perform this work. Two of us
(I.K. and T.T.) were partly supported by the Hungarian National
Science Fund OTKA, grant No.~2090.
\bigskip\medskip

\def\EPL#1#2{{\it Europhys.~Lett.}~{\bf#1#2}}
\def\JPL#1#2{{\it J.~Physique Lett.}~{\bf#1#2}}
\def\JPA#1#2{{\it J.~Phys.~A: Math.~Gen.}{\bf#1#2}}
\def\PRB#1#2{{\it Phys.~Rev.}~{\bf B}{\bf#1#2}}
\def\PRL#1#2{{\it Phys.~Rev.~Lett.}~{\bf#1#2}}
\def\JPI#1#2{{\it J.~Phys.~I.~France}~{\bf#1#2}}

\noindent{\bf References}
\bigskip

\noindent[1] M\'ezard M, Parisi G and Virasoro M 1987 {\it Spin Glass
Theory and Beyond} (Singapore: World Scientific)

\noindent[2] M\'ezard M and Parisi G 1990 \JPA 23 L1229

\noindent[3] M\'ezard M and Parisi G 1991 \JPI 1{} 809

\noindent[4] M\'ezard M and Parisi G 1992 \JPI 2{} 2231


\noindent[5] de Almeida J R L and Bruinsma R 1983 \PRB 35 7267

\noindent[6] M\'ezard M and Young A P 1992 \EPL 18 653

\noindent[7] Garel T and Orland H 1988 \EPL 6{} 307

\noindent[8] Shaknovich E I and Gutin A M 1989 \JPA 22 1647

\noindent[9] Sasai M and Wolynes P G 1990 \PRL 65 2740

\noindent[10] Bouchaud J-P, M\'ezard M and Yedidia J S 1991 \PRL
67 3840

\noindent[11] Parisi G 1980 \JPA 13 1887

\noindent[12] De Dominicis C, Kondor I and Temesv\'ari T 1994
to be published in \JPI {}{}

\noindent[13] Kondor I and De Dominicis C 1986 \EPL 2{} 617

\noindent[14] De Dominicis C and Kondor I 1984 \JPL 45 L-205

\noindent[15] De Dominicis C and Kondor I 1985 \JPL 46 L-1037
\bigskip\medskip
{\bf \noindent Appendix }
\bigskip
Our purpose here is to sketch the derivation of the replicon contributions
to Eqs. (65-76) through what we called the representative vector.

Eq.~(58), written out in the original, Cartesian coordinates, reads:

$$M_{\a\b,\c\d }=\Sum{ij}\bra{\a\b}i\rangle \tilde M_{ij}
\widetilde{\bra{j}}\c\d\rangle \eqno (A.1)$$
where $\ket i$ now means any of the ${1\over 2}n(n-1)$
new basis vectors, and $\widetilde{\ket j}$ are their
biorthogonal
counterparts.

We are interested here in the contribution of the
replicon family to
(A.1) only, i.e.~in the partial sum, to be denoted by
$M_{\a\b,\c\d }^R$, where $i$ and $j$ are restricted
to the replicon sector. But $\tilde M$ is diagonal
in that sector, $\tilde M_{ij}=\lambda_i \Kr ij $,
so we have
$$M_{\a\b,\c\d }^R=\Sum{i\in R}\bra{\a\b}i\rangle \lambda_i
\widetilde{\bra i}\c\d\rangle .\eqno (A.2)$$
Now let $\a\cap\b=r$. Then the replicon vectors $\ket i$
contributing to (A.2) have to be such
that they have nonzero components on the $r^{\rm th}$
level of the Parisi hierarchy. From
the description of these vectors given in the main text
we know, however, that
their components on every other level are then identically
zero, and furthermore
even on the $r^{\rm th}$ level their nonzero components
are concentrated inside a single
block of size $p_r\times p_r$.
Evidently, the component $(\a\b)$ must belong to this block.
Although
we have not actually determined the biorthogonal set
$\widetilde{\bra i}$ (and our purpose here
is to show that we do not need to, either),
it is obvious that $\widetilde{\bra i}$ will share
the above properties of $\ket i$: it will  have nonzero
components on the $r^{\rm th}$ level,
and there inside the same $p_r\times p_r$ block only.
It follows that $(\c\d)$ must belong to the
same block and $\a\cap\b=\c\cap\d=r$.

With this we have identified the set of replicon
vectors that, for a given $\a,\b,\c,\d$,
can give a nonzero contribution to $M_{\a\b,\c\d}^R$.
This set can be decomposed into orthogonal
classes, labelled by the triplet of integers $r,k,l$
($k,l\ge r+1$), as explained in Sec.~3. In a
given class ($r,k,l$) we still have several nonorthogonal
replicon vectors and, in principle,
they all contribute to (A.2). What we wish to show
is that, in fact, one
can choose a single vector from each class ($r,k,l$)
in such a way as to exhaust the contribution
of the whole class. We have called this vector the
representative vector of the
class.

The choice of the representative vector is not unique.
It is best
to choose it such that the component ($\c\d$)
belong to the "darkest" block, where
the vector has the components $A$. (Consult Figs.~5,6,7.)
Now, as we have already
pointed out in Sec.~3, the subspace orthogonal
to the representative vector thus
chosen is spanned by vectors that have zero components
over this "$A$-block". In
particular, if $\ket{r;k,l}$ is the representative vector
of the class ($r,k,l$) then the biorthogonal
counterparts of all the other members of the class
will lie in the space orthogonal
to $\ket{r;k,l}$, hence their components in the
"$A$-block" where also the pair ($\c\d$) resides
must necessarily be zero and thus their scalar
product with the unit vector $\ket{\c\d}$
will vanish. Therefore, the only contribution from the class
($r,k,l$) comes from the
representative vector, indeed.

The summation in (A.2) then runs over the representative
vectors only, so we
can rewrite (A.2) as
$$M_{\a\b,\c\d}^R=\sum\limits^{R+1}_{k=r+1}\sum\limits^{R+1}
_{l=r+1} \lambda(r;k,l) \bra{\a\b}r;k,l\rangle \widetilde
{\bra{r;k,l}}\c\d\rangle .\eqno (A.3)$$
We now decompose $\widetilde{\bra{r;k,l}}$ into
components parallel and orthogonal to $\ket{r;k,l}$.
The orthogonal
component will not contribute to (A.3) for the same reasons
as above. So we are
left with the parallel component only, which, in view of
$\widetilde{\bra{r;k,l}}r;k,l\rangle=1$ and of the
normalisation
of $\ket{r;k,l}$, is nothing but the representative
vector itself. So we can finally write
$$M_{\a\b,\c\d}^R=\sum\limits^{R+1}_{k=r+1}\sum\limits^{R+1}
_{l=r+1} \lambda(r;k,l) \bra{\a\b}r;k,l\rangle
\bra{r;k,l}\c\d\rangle .\eqno (A.4)$$

Although the above consideration is quite trivial
really, one may find it
mystifying that it is possible to reconstruct a matrix
from selecting a single
vector from each class of its eigenvectors which
are nonorthogonal within the
class. As it transpires from the proof, the key factor
is that the subspace
orthogonal to the representative vector is composed
of vectors having vanishing
components over the "$A$-block", and this in turn
hinges upon the common property
of all replicons, namely that the sum of their components
in each row is zero.
A little more reflection will show one, however,
that the underlying reason is
that the matrix $M_{\a\b,\c\d}^R$ does not depend on
$\a$, $\b$, $\c$ and $\d$ separately, only on the various overlaps
formed out of these indices, therefore the puzzling
property of the representative
vectors carrying all the information about $M^R$ can be
directly linked to the
ultrametric symmetries of $M$.

As an illustration of the use of (A.4), let us calculate the
diagonal components $M_{\a\b,\a\b}^R$. There will be three kinds
of terms contributing to (A.4):

(i) $k=l=r+1$:
$$\bra{r;r+1.r+1}\a\b\rangle^2=A^2={p_r-3p_{r+1}\over p_{r+1}
^2 (p_r-p_{r+1})}={2\mu(r;r+1,r+1)\over n\d_r}\quad ,\eqno
(A.5)$$
where use has been made of Eqs.~(32), (33).

(ii) $k\ge r+2$, $l=r+1$:
$$\bra{r;k,r+1}\a\b\rangle^2=A^2={p_r-2p_{r+1}\over
p_{r+1}(p_r-p_{r+1})}\left({1\over p_k}-{1\over p_{k-1}}
\right)={2\mu(r;k,r+1)\over n\d_r}\quad,\eqno (A.6)$$
where we have used (34) and (35).

(iii) $k\ge r+2$, $l\ge r+2$:
$$\bra {r;k,l}\a\b\rangle^2=A^2=\left({1\over p_k}-{1\over p_{k-1}}
\right)\left({1\over p_l}-{1\over p_{l-1}}
\right)={2\mu(r;k,l)\over n\d_r}\quad,\eqno (A.7)$$
see (36), (37).

Substituting (A.5), (A.6) and (A.7) back into (A.4) we find
$$M_{\a\b,\a\b}^R=\left(M^R\right)^{r,r}_{R+1,R+1}=
\sum\limits^{R+1}_{k=r+1}\sum\limits^{R+1}_{l=r+1}
{2\mu(r;k,l)\over n\d_r}\lambda (r;k,l)$$
which is precisely the second (replicon) term quoted in
(65a). The replicon contributions to (66a), (67a), (74a), (75a)
and (76a) can be worked out similarly.
\bye